\documentstyle[aps]{revtex}

\input epsf.sty

\preprint{UCTP-103-99}

\begin{document}

\title{Theory of the Magnetic Catalysis of Chiral Symmetry
Breaking in QED}

\author{V.P.~Gusynin$^{1}$, V.A.~Miransky$^{1}$,
and I.A.~Shovkovy$^{2}$\thanks{On leave of absence from 
Bogolyubov Institute for Theoretical Physics, 252143, Kiev,
Ukraine.}}

\address{$^{1}$ Bogolyubov Institute for Theoretical Physics,
252143, Kiev, Ukraine\\
%$^{2}$Department of Physics, Nagoya University,
%Nagoya 464-8602, Japan \\
$^{2}$Physics Department, University of Cincinnati,
Cincinnati, Ohio 45221-0011}

\date{\today}
\draft

\maketitle

\begin{abstract}
The theory of the magnetic catalysis of chiral symmetry breaking
in QED is developed. An approximation for the Schwinger-Dyson
equations describing reliably this phenomenon is established,
{\em i.e.}, it is shown that there exists a consistent truncation
of those equations in this problem. The equations are solved both
analytically and numerically, and the dynamical mass of fermions
is determined. 
\end{abstract}


\section{Introduction}

Recently the magnetic catalysis of dynamical chiral symmetry
breaking has been established as a universal phenomenon in 2+1
and 3+1 dimensions: a constant magnetic field leads to the
generation of a fermion dynamical mass even at the weakest
attractive interaction between fermions \cite{1,2,3}. The essence
of this effect is the dimensional reduction $D\to D-2$ in the
dynamics of fermion pairing in a magnetic field: at weak
coupling, this dynamics is dominated by the lowest Landau level
(LLL) which is essentially ($D-2$)-dimensional \cite{1,2,3}. The
effect may have interesting applications in condensed matter
physics \cite{cond-mat} and cosmology \cite{1,4,5,6,FI-cosm}.

In particular, this phenomenon was considered in 3+1 dimensional
QED \cite{2,3,4,5,6,7,8,9,15,20,10,11}. Since the dynamics of the
LLL is long-range (infrared), and the QED coupling constant is 
weak in the infrared region, one may think that the rainbow
(ladder) approximation is reliable in this problem. As was shown
in Refs.~\cite{2,3,5}, in all the covariant gauges, the dynamical
mass of fermions in this approximation is
\begin{equation}
m_{dyn} = C \sqrt{|eB|}
\exp\left[-\frac{\pi}{2}
\left(\frac{\pi}{2\alpha}\right)^{1/2}
\right],
\label{m_dyn}
\end{equation}
where $B$ is a magnetic field, the constant $C$ is of order one
and $\alpha$ is the renormalized coupling constant related to the
scale $\mu^2\sim |eB|$.

Are higher order contributions indeed suppressed in this problem?
The answer is ``no". As was shown in Refs.~\cite{3,4}, because of
the (1+1)-dimensional form of the fermion propagator of the LLL
fermions, there are relevant higher order contributions. In
particular, considering this problem in the improved rainbow
approximation (with the bare vertex in the Schwinger-Dyson
equations for both the fermion propagator and the polarization
operator), it was shown that, in all the covariant gauges, the
fermion mass $m_{dyn}$ is given by Eq.~(\ref{m_dyn}) but with
$\alpha\to\alpha/2$ \cite{3}.

As we wrote in the paper \cite{3}, ``it is a challenge to define
the class of all those diagrams in QED in a magnetic field that
give a relevant contribution in this problem". In this paper, we
will solve the problem. (A brief outline of our results was given
in Ref.~\cite{12}.) We will show that there exists a
(non-covariant) gauge in which the Schwinger-Dyson equations
written in the improved rainbow approximation are reliable: in
other words, in that gauge, there exists a consistent truncation
of the Schwinger-Dyson equations for this {\em non-perturbative}
problem. The expression for $m_{dyn}$ takes the following form,
\begin{equation}
m_{dyn} =\tilde C \sqrt{|eB|} F(\alpha)\exp\left[-\frac{\pi}
{\alpha\ln\left(C_1/N\alpha\right)}\right], \label{m}
\end{equation}
where $N$ is the number of fermion flavors,$F(\alpha) \simeq
(N\alpha)^{1/3}$, $C_1\simeq 1.82\pm 0.06$ and the constant
$\tilde C$ is of order one.

This expression for $m_{dyn}$ is essentially different from that
in the rainbow approximation (\ref{m_dyn}). As we will see, this
reflects rather rich and sophisticated dynamics in this problem.

The paper is organized as follows. In Section \ref{secII} the
Schwinger-Dyson equations in QED in a magnetic field are
discussed. In Section \ref{secIII} we define a (non-covariant)
gauge in which the improved rainbow approximation for these
equations is reliable. In Section \ref{secIV} the loop expansion
for the Schwinger-Dyson equations is considered and it is
confirmed that the improved rainbow approximation is indeed
reliable in the gauge introduced in Section \ref{secIII}. In
Section \ref{secV} the Schwinger-Dyson equations in this
approximation are solved both analytically and numerically. In
Section \ref{secVI} we summarize the main results of the paper.
The details of our analysis and some useful formulas and
relations are presented in Appendices \ref{secA}, \ref{secB} 
and \ref{secC}.

\section{The Schwinger-Dyson equations in QED in a magnetic
field.} 
\label{secII}

The Lagrangian density of massless QED in a magnetic field is
\begin{equation}
{\cal L}=-\frac{1}{4}F^{\mu\nu}F_{\mu\nu}+\frac{1}{2}
\left[\bar{\psi}, (i\gamma^{\mu}D_{\mu})\psi\right]+J^\mu A_\mu,
\label{lag}
\end{equation}
where the covariant derivative $D_{\mu}$ is
\begin{equation}
D_{\mu}=\partial_{\mu}-i e A_{\mu}
\label{covderiv}
\end{equation}
and the source $J^\mu$ provides a chosen external field:
$\langle0|A_\mu|0\rangle=A_\mu^{ext}$. As we will see below, in
the case of a constant magnetic field $B$, with
\begin{equation}
A_\mu^{ext}=\left(0,-{B\over2}x_2,{B\over2}x_1,0\right),
\label{extpotential}
\end{equation}
the source $J^\mu=0$. Notice that the vector potential
$A_\mu^{ext}$ (\ref{extpotential}) corresponds to the so called
symmetric gauge for the external vector potential, and the
magnetic field is in the $+x_3$ direction.

Besides the Dirac index (n), the fermion field carries an
additional flavor index $a=1,2,\dots,N$. Then the Lagrangian
density in Eq.~(\ref{lag}) is invariant under the chiral
$SU_{L}(N)\times SU_{R}(N)\times U_V(1)$ symmetry (we will
discuss the anomalous $U_A(1)$ in Sec.~\ref{secVI}).

The Schwinger-Dyson (SD) equations in QED in external fields
were derived by Schwinger and Fradkin (for a review, see
Ref.~\cite{13}). The equations for the fermion propagator
$G(x,y)$ are
\begin{eqnarray}
&&G^{-1}(x,y)=S^{-1}(x,y)+\Sigma(x,y),\label{G{-1}}\\
&&\Sigma(x,y)=4\pi\alpha\gamma^\mu\int G(x,z)
\Gamma^\nu (z,y,z^\prime){\cal D}_{\nu\mu}
(z^\prime,x)d^4zd^4z^\prime.
\label{SD-fer}
\end{eqnarray}
Here $S(x,y)$ is the bare fermion propagator in the external
field $A_{\mu}^{ext}$, $\Sigma(x,y)$ is the fermion mass
operator, and ${\cal D}_{\mu\nu}(x,y)$, $\Gamma^\nu (x,y,z)$ are
the full photon propagator and the full amputated vertex.

The full photon propagator satisfies the equations
\begin{eqnarray}
{\cal D}^{-1}_{\mu\nu}(x,y)&=&D^{-1}_{\mu\nu}(x-y)
+\Pi_{\mu\nu}(x,y), \label{SD-pho}\\
\Pi_{\mu\nu}(x,y)&=&-4\pi\alpha \mbox{tr} \gamma_{\mu}
\int d^4 u d^4 z G(x,u)\Gamma_{\nu} (u,z,y) G(z,x),
\label{Pi_munu}
\end{eqnarray}
where $D_{\mu\nu}(x-y)$ is the free photon propagator and
$\Pi_{\mu\nu}(x,y)$ is the polarization operator.

The equation for the external photon field
$A_\mu^{ext}=\langle0|A_\mu|0\rangle$ in a covariant gauge is
\begin{equation}
\Box A_\mu^{ext}-\lambda\partial_\mu\partial^\nu A_\nu^{ext}=
-J_\mu-\langle0|j_\mu|0\rangle,
\end{equation}
where the vacuum current $\langle0|j_\mu|0\rangle$ is
$\langle0|j_\mu|0\rangle=-e{\rm tr}(\gamma_\mu G(x,x))$ and
$\lambda$ is the gauge parameter. Notice that for $A_\mu^{ext}$
in Eq.~(\ref{extpotential}), corresponding to a constant magnetic
field, the source $J_\mu$ is zero. Indeed, in this case,
$J_\mu=-\langle0|j_\mu|0\rangle$, and
$\langle0|j_\mu|0\rangle=-e{\rm tr}(\gamma_\mu G(x,x))=0$ because
of the symmetry $SO(2)\times SO(1,1)$, with $SO(2)$ and $SO(1,1)$
corresponding to rotations in the $x_1-x_2$-plane and Lorentz
transformations in the $x_0-x_3$-hyperplane, respectively.

The bare fermion propagator $S(x,y)$ in a constant magnetic field
was calculated by Schwinger \cite{14}. In the symmetric gauge
(\ref{extpotential}), it has the form 
\begin{equation}
S(x,y)=\exp\left(ie x^{\mu} A^{ext}_{\mu}(y)\right)
\tilde{S}(x-y),
\end{equation}
where the Fourier transform of $\tilde S$ is
\begin{eqnarray}
\tilde{S}(p) &=&\int\limits^\infty_0 ds 
\exp \left[is\left(p^2_0-p^2_3
-p^2_{\perp}\frac{\tan(eBs)}{eBs}-m\right)\right] \nonumber \\
&\cdot &
\left[(p^0\gamma^0-p^3\gamma^3+m)(1+\gamma^1\gamma^2\tan(eBs))
-p_{\perp}\gamma_{\perp}(1+\tan^2(eBs))\right].
\label{SFourier}
\end{eqnarray}
Here a transverse vector $p_\perp=(p^1,p^2)$ and $m$ is a
fermion mass. Then, using the identity
\begin{equation}
ix^\mu A_\mu^{ext}(z)+iz^\mu A_\mu^{ext}(y)
=ix^\mu A_\mu^{ext}(y)+i(x-y)^\mu A_\mu^{ext}(z-y)
\end{equation}
for the vector potential (\ref{extpotential}), it is not
difficult to show directly from the SD equations (\ref{G{-1}}),
(\ref{SD-fer}), (\ref{SD-pho}) and (\ref{Pi_munu}) that
\begin{mathletters}
\begin{eqnarray}
&&G(x,y)=\exp\left(ie x^{\mu} A^{ext}_{\mu}(y)\right)
\tilde{G}(x-y),
\label{14a}\\
&&\Gamma(x,y,z)=\exp\left(ie x^{\mu} A^{ext}_{\mu}(y)\right)
\tilde{\Gamma}(x-z,y-z),
\label{14b}\\
&&{\cal D}_{\mu\nu}(x,y)=\tilde{{\cal D}}_{\mu\nu}(x-y),
\label{14c}\\
&&\Pi_{\mu\nu}(x,y)=\tilde{\Pi}_{\mu\nu}(x-y).
\label{14d}
\end{eqnarray}
\label{14}
\end{mathletters}
In other words, in a constant magnetic field, the Schwinger phase
is universal for Green functions containing one fermion field,
one antifermion field, and any number of photon fields, and the
full photon propagator is translation invariant.

Our aim is to show that there exists a gauge in which the
approximation with a bare vertex,
\begin{equation}
\Gamma^{\mu}(x,y,z)=\gamma^{\mu} \delta(x-y)\delta(x-z),
\label{bare-ver}
\end{equation}
is reliable for the description of spontaneous chiral symmetry
breaking in a magnetic field.

\section{Non-covariant gauge and the improved rainbow
approximation for the SD equations}
\label{secIII}

In this section, we will show that there is a (non-covariant)
gauge in which the approximation with a bare vertex
(\ref{bare-ver}) (the improved rainbow approximation) is reliable.

We begin  by recalling the following facts concerning the problem
of the magnetic catalysis of chiral symmetry breaking \cite{3}:

\begin{enumerate}
\item At weak coupling, there is the LLL dominance in the
dynamics of fermion pairing. It is because of the presence of the
large Landau gap of order $\sqrt{|eB|}$, which is much larger
than the dynamical fermion mass $m_{dyn}$ (for weak coupling). In
other words, higher Landau levels decouple from the infrared
dynamics (with $k\ll \sqrt{|eB|}$)  in the same way as it happens
with Kaluza-Klein (KK) modes in KK theories of gravity. This fact
was explicitly shown in the Nambu-Jona-Lasinio model \cite{3} and
in QED \cite{15}.

\item The propagator $\tilde{S}(p)$ (\ref{SFourier}) can be
expanded over the Landau levels \cite{16,3}. The contribution
from the LLL is
\begin{equation}
\tilde{S}_{\rm LLL}(p)=2i e^{-(p_{\perp}l)^2}\frac{\hat{p}_{\parallel}+m}
{p^2_{\parallel}-m^2} O^{(-)},
\label{tildeS}
\end{equation}
where the magnetic length $l=|eB|^{-1/2}$, $p_{\perp}=(p^1,p^2)$,
$p_{\parallel}=(p^0,p^3)$, and $\hat{p}_{\parallel} =p^0
\gamma^0-p^3 \gamma^3$. The matrix $O^{(-)}\equiv
\left(1-i\gamma^{1}\gamma^{2} {\rm sign}(eB)\right)/2$ is the
projection operator on the fermion states with the spin polarized
along the magnetic field. This point and Eq.~(\ref{tildeS})
clearly reflect the (1+1)-dimensional character of the dynamics
of fermions in the LLL. This property is preserved also in the
case when the fermion mass is generated dynamically ($m=m_{\rm
dyn}$) \footnote{As was shown in Refs.\cite{3,Elias}, despite the
dimensional reduction $3+1\rightarrow 1+1$ in the fermion
propagator in a magnetic field, there is a genuine spontaneous
chiral symmetry breaking in this problem: the
Mermin-Wagner-Coleman (MWC) theorem \cite{MWC}, forbidding
spontaneous breakdown of continuous symmetries at $D=1+1$, is not
applicable to this case. The point is that the MWC theorem is
based on the fact that gapless Nambu-Goldstone (NG) bosons cannot
exist in $1+1$ dimensions. On the other hand, since the NG bosons
connected with spontaneous chiral symmetry breaking are neutral,
there is no dimensional reduction of their propagator in a
magnetic field \cite{3,Elias}. Therefore their propagator is
($3+1$)-dimensional and there are no obstacles for their
existence in this case.}.

\item The presence of the projection operator $O^{(-)}$ implies
that the bare vertex for fermions from the LLL is
$O^{(-)}\gamma^\mu O^{(-)} =O^{(-)}\gamma_{\parallel}^\mu $.
Therefore the LLL fermions couple only to the longitudinal
$(0,3)$ components of the photon field.

\item In the one-loop approximation, with fermions from the LLL,
the photon propagator takes the following form in covariant
gauges \cite{3},
\begin{eqnarray}
{\cal D}_{\mu\nu}(q)=-i\left[\frac{1}{q^2}g^{\perp}_{\mu\nu}
+\frac{q^{\parallel}_{\mu} 
q^{\parallel}_{\nu}}{q^2 q_{\parallel}^2}
+\frac{1}{q^2+q_{\parallel}^2 \Pi (q_{\perp}^2,q_{\parallel}^2)}
\left(g^{\parallel}_{\mu\nu}-\frac{q^{\parallel}_{\mu}
q^{\parallel}_{\nu}}{q_{\parallel}^2}\right)
-\frac{\lambda}{q^2}\frac{q_{\mu}q_{\nu}}{q^2}\right],
\label{cov-gauge}
\end{eqnarray}
where the symbols $\perp$ and $\parallel$ in $g_{\mu\nu}$ are
related to the $(1,2)$ and $(0,3)$ components, respectively, and
$\lambda$ is a gauge parameter. The explicit expression for $\Pi
(q_{\perp}^2, q_{\parallel}^2)=\exp[-(q_{\perp} l)^2/2] \Pi
(q_{\parallel}^2)$ is given in Refs.~\cite{17,3}. For our
purposes, it is sufficient to know its asymptotes,
\begin{eqnarray}
\Pi (q_{\parallel}^2)\simeq \frac{\bar{\alpha}}{3\pi}
\frac{|eB|}{m_{dyn}^2}, \quad \mbox{as}
\quad |q_{\parallel}^2| \ll m_{dyn}^2,
\label{Pi-IR}\\
\Pi (q_{\parallel}^2)\simeq -\frac{2\bar{\alpha}}{\pi}
\frac{|eB|}{q_{\parallel}^2} \quad \mbox{as}
\quad |q_{\parallel}^2| \gg m_{dyn}^2,
\label{Pi-UV}
\end{eqnarray}
where $\bar{\alpha}=N\alpha$.

Notice that the polarization effects are absent in the transverse
components of ${\cal D}_{\mu\nu}(q)$. This is because, as was
already pointed out in item 3 above, fermions from the LLL couple
only to the longitudinal components of the photon field.

\item Then, there is a strong screening effect in the
$\left(g^{\parallel}_{\mu\nu}-q^{\parallel}_{\mu}
q^{\parallel}_{\nu}/q_{\parallel}^2\right)$ component of the
photon propagator. Eq.~(\ref{Pi-UV}) implies that
\begin{equation}
\frac{1}{q^2+q_{\parallel}^2 \Pi (q_{\perp}^2, q_{\parallel}^2)}
\simeq \frac{1}{q^2-M_{\gamma}^2},
\label{screenedphoton}
\end{equation}
with
\begin{equation}
M_{\gamma}^2= \frac{2\bar{\alpha}}{\pi}|eB|
\label{M_gamma}
\end{equation}
for $m^2\ll |q_{\parallel}^2| \ll |eB|$ and $|q_{\perp}^2| \ll
|eB|$. This is reminiscent of the Higgs effect in the
(1+1)-dimensional QED (Schwinger model) \cite{18,19}.
\end{enumerate}

We emphasize that the infrared dynamics in this problem is very
different from that in the Schwinger model: since photon is
neutral, there is no dimensional reduction for its field in a
magnetic field, and there is the four-dimensional
$q^2=q_{\parallel}^2-q_{\perp}^2$ in the photon propagator
(\ref{screenedphoton}). However, the tensor and the spinor
structure of this dynamics are exactly the same as in the
Schwinger model: indeed, the LLL fermion propagator
(\ref{tildeS}) and the vertex $O^{(-)}\gamma^\mu
O^{(-)}=O^{(-)}\gamma_{\parallel}^\mu$ are two-dimensional, and
only the longitudinal $(0,3)$ components of a photon field are
relevant here. This point will be crucial for finding a gauge in
which the improved rainbow approximation (with the bare vertex
(\ref{bare-ver})) is reliable\footnote{Since an external magnetic
field does not lead to confinement of fermions, their mass is
gauge invariant in QED in a magnetic field. Therefore {\it any}
gauge can be used for the calculations of the mass {\it if}
either the calculations provide the exact result or a good
approximation is used: {\em i.e.}, one can show that corrections
to the obtained result are small. Below we will define such a
gauge in this model.}.

We recall that, as was shown in Ref.~\cite{3}, despite the
smallness of $\alpha$, the expansion in $\alpha$ is broken in
covariant gauges in this problem. The reason is that, because of
the smallness of $m_{dyn}$ in Eq.~(\ref{m_dyn}) as compared to
$\sqrt{|eB|}$, there are mass singularities,
$\ln(|eB|/m_{dyn}^2)\sim\alpha^{-1/2}$, in infrared dynamics. In
Appendix~\ref{secA}, we analyze these singularities in the
vertex. Calculating the one-loop correction to the vertex,  one
finds that, when external momenta are of order $m_{dyn}$ or less,
there are contributions of order $\alpha\ln^2(|eB|/m_{dyn}^2)\sim
O(1)$. They come from the term $q_{\mu}^{\parallel}
q_{\nu}^{\parallel}/q^2 q_{\parallel}^2$ in ${\cal
D}_{\mu\nu}(q)$ in Eq.~(\ref{cov-gauge}).

How can one avoid such mass singularities? A solution is
suggested by the Schwinger model. It is known that there is a
gauge in which the full vertex is just the bare one \cite{19}. It
is the gauge with a bare photon propagator
\begin{equation}
D_{\alpha\beta}(k)=-i\frac{1}{k^2}\left(g_{\alpha\beta}-\frac
{k_{\alpha} k_{\beta}}{k^2}\right)
-i d(k^2) \frac{k_{\alpha} k_{\beta}}{(k^2)^2}
\end{equation}
with the (non-local) gauge function $d= 1/(1+\Pi)$,
where the polarization function $\Pi(k^2)=-e^2/\pi k^2$ in the
Schwinger model (of course, here $\alpha,~\beta =0,1$). Then,
the full propagator is proportional to $g_{\alpha\beta}$,
\begin{eqnarray}
{\cal D}_{\alpha\beta}(k)&=&D_{\alpha\beta}(k)
+ i\left(g_{\alpha\beta}- \frac{k_{\alpha} 
k_{\beta}}{k^2} \right)
\frac{\Pi(k^2)}{k^2(1+\Pi(k^2))}
=-i \frac{g_{\alpha\beta}}{k^2(1+\Pi(k^2))}.
\label{gauge-S}
\end{eqnarray}
The point is that since now ${\cal D}_{\alpha\beta}(k)\sim
g_{\alpha\beta}$ and since the fermion mass $m=0$ in the
Schwinger model, all loop contributions to the vertex are
proportional to 
\begin{equation}
P_{2n+1}\equiv  \gamma_{\alpha} \gamma_{\lambda_{1}} \dots
\gamma_{\lambda_{2n+1}} \gamma^{\alpha}=0
\end{equation}
in that gauge and,
therefore, disappear\footnote{$P_{2n+1}=0$ follows from the two
identities for the two-dimensional Dirac matrices:
$\gamma_{\alpha} \gamma_{\lambda} \gamma^{\alpha}=0$ and
$\gamma_{\lambda_{i}} \gamma_{\lambda_{i+1}}=g_{\lambda_{i}
\lambda_{i+1}} +\varepsilon_{\lambda_{i} \lambda_{i+1}}
\gamma_{5}$ ($\gamma_{5}=\gamma_{0} \gamma_{1}$,
$\varepsilon_{\alpha\beta} =-\varepsilon_{\beta\alpha}$,
$\varepsilon_{01}=1$).}.

Let us return to the present problem. As it was emphasized above,
the tensor and the spinor structure of the LLL dynamics is
(1+1)-dimensional. Now, take the bare propagator
\begin{equation}
D_{\mu\nu}(q)=-i\frac{1}{q^2}\left(g_{\mu\nu}
-\frac{q_{\mu} q_{\nu}} {q^2}\right)
-i d(q_{\perp}^2, q_{\parallel}^2) \frac{q^{\parallel}_{\mu}
q^{\parallel}_{\nu}}{q^2 q_{\parallel}^2}
\end{equation}
with $d =-q_{\parallel}^2\Pi/[q^2+q_{\parallel}^2\Pi] +
q_{\parallel}^2/q^2$. Then, the full propagator is
\begin{eqnarray}
{\cal D}_{\mu\nu}(q)&=&D_{\mu\nu}(q)
+ i\left(g^{\parallel}_{\mu\nu}-
\frac{q^{\parallel}_{\mu} q^{\parallel}_{\nu}}{q_{\parallel}^2}
\right) \frac{q_{\parallel}^2 \Pi(q_{\perp}^2, q_{\parallel}^2)}
{q^2[q^2+q_{\parallel}^2 \Pi(q_{\perp}^2, q_{\parallel}^2)]}
\nonumber \\
&=&-i\frac{g^{\parallel}_{\mu\nu}}{q^2+q_{\parallel}^2
\Pi(q_{\perp}^2, q_{\parallel}^2)} -i
\frac{g^{\perp}_{\mu\nu}}{q^2}
+i\frac{q^{\perp}_{\mu}q^{\perp}_{\nu} + q^{\perp}_{\mu}
q^{\parallel}_{\nu} + q^{\parallel}_{\mu}q^{\perp}_{\nu}}
{(q^2)^2}.
\label{non-l}
\end{eqnarray}
The crucial point is that, as was pointed out above, the
transverse degrees of freedom decouple from the LLL dynamics.
Therefore only the first term in ${\cal D}_{\mu\nu}(q)$,
proportional to $g^{\parallel}_{\mu\nu}$, is relevant.

Notice now that dangerous mass singularities in loop corrections
to the vertex might potentially occur only in the terms
containing $\hat{q}^{\parallel}_i =q_{i}^{0} \gamma^{0}-q_{i}^{3}
\gamma^{3}$ from a numerator ($\hat{q}^{\parallel}_i+m_{dyn}$) of
{\it each} fermion propagator in a diagram (all other terms
contain positive powers of $m_{dyn}$, coming from at least some 
of the numerators and, therefore, are harmless). However, because
of the same reasons as in the gauge (\ref{gauge-S}) in the
Schwinger model, all those potentially dangerous terms disappear
in the gauge (\ref{non-l}). Therefore all the loop corrections to
the vertex are suppressed by positive powers of $\alpha$ in this
gauge. This in turn implies that those loop corrections may
result only in a change $\tilde C\sim O(1)\rightarrow \tilde
C^\prime\sim O(1)$ in expression (\ref{m}.) In other words, in
gauge (\ref{non-l}) there exists a {\it consistent} truncation of
the SD equations and the problem is essentially soluble in that
gauge. \footnote{The gauge (\ref{non-l}) is unique in that. In
other gauges, there is an infinite set of diagrams giving relevant
contributions to the vertex. Therefore, in other gauges, one
needs to sum up an infinite set of diagrams to recover the same
result for the fermion mass.}

In the next section, we will consider the loop expansion for the
SD equations in this problem in more detail.

\section{The loop expansion for the SD equations}
\label{secIV}

The consideration of mass singularities in loop corrections given
at the end of the previous section was, though general, somewhat
heuristic. First of all, one has to define more rigorously the
perturbative expansion for the SD equations which is used in this
problem. It is the loop expansion based on the CJT (Cornwall,
Jackiw, and Tomboulis) effective action $\Gamma(G,{\cal
D}_{\mu\nu})$ for composite operators \cite{21} (for a
review see Ref.~\cite{22}).
The conditions for extrema of $\Gamma$ yield the
SD equations,
\begin{equation}
\frac{\delta\Gamma}{\delta G(x,y)}=0, 
\quad \frac{\delta\Gamma}{\delta
{\cal D}_{\mu\nu}(x,y)}=0.
\end{equation}
In the loop expansion for $\Gamma$, the {\it full} photon and
fermion propagators are used in two-particle irreducible diagrams
for $\Gamma$. In QED, the problem is essentially reduced to the
loop expansion (with the full photon and fermion propagators) for
the vertex.

The full photon propagator is given by Eq.~(\ref{non-l}), and the
full propagator for fermions from the LLL has the form
\begin{equation}
\tilde{G}(p)=2 i e^{-(p_{\perp}l)^2} \frac{ A(p_{\parallel}^2)
\hat{p}_{\parallel} +B(p_{\parallel}^2) }
{A^2(p_{\parallel}^2) p^2_{\parallel} - B^2(p_{\parallel}^2)}
O^{(-)}
\label{gen-sol}
\end{equation}
(compare with Eq.~(\ref{tildeS}) and see below). Here
$B(p_{\parallel}^2)$ is a dynamical mass function of fermions.

In this section we will derive the SD equations for the fermion
propagator in the one-loop and two-loop approximations. It will
be shown, that while the one-loop approximation, coinciding with
the improved rainbow approximation, is reliable in non-covariant
gauge (\ref{non-l}), it is not reliable in covariant gauges
(\ref{cov-gauge}).

\begin{figure}
\epsfbox{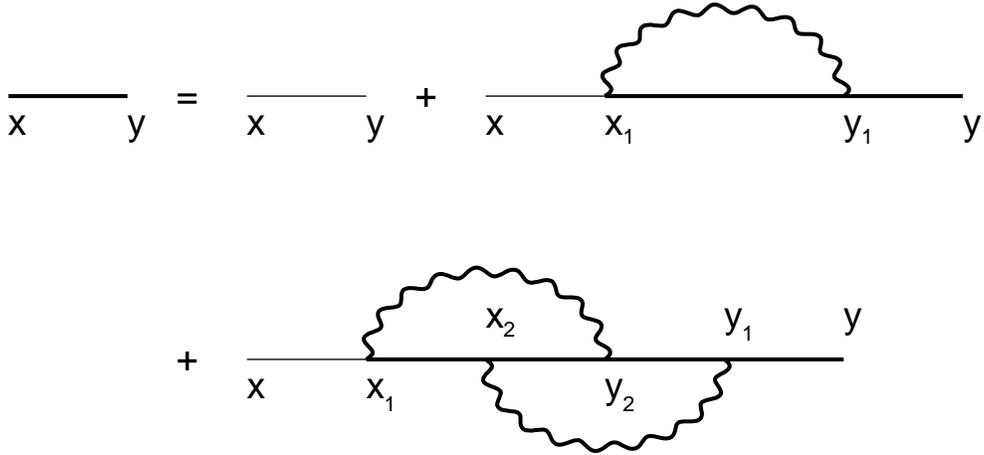}
\caption{The diagrammatic representation of the SD equation in
coordinate space at two-loop order. A thin solid line corresponds
to the bare fermion propagator; a bold line corresponds to the full
fermion propagator, and a wavy line designs the full photon
propagator.}
\label{fig-SD}
\end{figure}

From Eqs.~(\ref{G{-1}}) and (\ref{SD-fer}) one gets the following
equation for the fermion propagator $G(x,y)$ in the two-loop
approximation (by using the vertex function in the one-loop
approximation, see Eq.~(\ref{14b}) and Eq.~(\ref{A3}) in 
Appendix~\ref{secA}),
\begin{eqnarray}
&&G(x,y)=S(x,y)-4\pi \alpha \int d^4x_1 d^4y_1 
S(x,x_1)\gamma^{\mu}
G(x_1,y_1)\gamma^{\nu}G(y_1,y){\cal D}_{\mu\nu}(x_1-y_1) 
\nonumber \\
&&+(4\pi \alpha)^2 \int d^4 x_1 d^4 x_2 d^4 y_1 d^4 y_2
S(x,x_1) \gamma^{\mu} G(x_1,x_2) \gamma^{\sigma} G(x_2,y_2)
\gamma^{\nu} G(y_2,y_1) \gamma^{\rho} G(y_1,y)
{\cal D}_{\mu\nu}(x_1-y_2) {\cal D}_{\sigma\rho}(x_2-y_1).
\label{B1}
\end{eqnarray}
Here $S(x,y)$ is the bare fermion propagator of massless fermions
($m=0$). The graphic form of this equation is shown in
Fig.~\ref{fig-SD}. After extracting the Schwinger phase factors 
in the full and bare fermion propagators [see Eq.~(\ref{14a})],
\begin{equation}
G(x,y) = e^{ix^{\nu}A_{\nu}^{ext}(y)} \tilde{G}(x-y), \quad
S(x,y)=e^{ix^{\nu}A_{\nu}^{ext}(y)} \tilde{S}(x-y),
\label{B2}
\end{equation}
Eq.~(\ref{B1}) reads as
\begin{eqnarray}
\tilde{G}(x)&=&\tilde{S}(x)
-4\pi \alpha \int d^4 x_1 d^4 y_1 e^{ix A(x_1)+ix_1 A(y_1)}
\tilde{S}(x-x_1)\gamma^{\mu}
\tilde{G}(x_1-y_1)\gamma^{\nu}
\tilde{G}(y_1) {\cal D}_{\mu\nu}(x_1-y_1)\nonumber\\
&+&(4\pi \alpha)^2 \int d^4 x_1 d^4 x_2 d^4 y_1 d^4 y_2
e^{ix A(x_1)+ix_1 A(x_2)+ix_2 A(y_2)+iy_2 A(y_1)}\nonumber\\
&&\times\tilde{S}(x-x_1)\gamma^{\mu}
\tilde{G}(x_1-x_2)\gamma^{\sigma}
\tilde{G}(x_2-y_2)\gamma^{\nu}
\tilde{G}(y_2-y_1)\gamma^{\rho}
\tilde{G}(y_1) {\cal D}_{\mu\nu}(x_1-y_2)
{\cal D}_{\sigma\rho}(x_2-y_1)
\label{B3}
\end{eqnarray}
where $A_\mu^{ext}$ is given in Eq.~(\ref{extpotential}) and
the shorthand $xA^{ext}(y)$ stands for $x^\mu A_\mu^{ext}(y)$.

First, let us show that the solution to the above equation,
$\tilde{G}(x)$, allows the factorization of the dependence on 
the parallel and perpendicular coordinates,
\begin{equation}
\tilde{G}(x)=\frac{i}{2\pi l^2}
\exp\left(-\frac{x_{\perp}^2}{4l^2}\right)
g\left(x_{\parallel}\right)O^{(-)}.
\label{B4}
\end{equation}
Notice that this form for $\tilde{G}(x)$ is suggested by a
similar expression for the bare propagator,
\begin{equation}
\tilde{S}(x)=\frac{i}{2\pi l^2}
\exp\left(-\frac{x_{\perp}^2}{4l^2}\right)
s\left(x_{\parallel}\right)O^{(-)},
\label{B5}
\end{equation}
with
\begin{equation}
s\left(x_{\parallel}\right)
=\int\frac{d^2 k_{\parallel}}{(2\pi)^2}
e^{-ik_{\parallel}x_{\parallel}}
\frac{\hat{k}_{\parallel}+m}{k_{\parallel}^2-m^2}
\label{B6}
\end{equation}
(see Eq.~(\ref{tildeS}); in the chiral limit, as in the present
problem, the bare mass $m=0$). In order to perform the
integrations over the perpendicular components of $x_1$ and $y_1$
in Eq.~(\ref{B3}), it is convenient to make use of the photon
propagator in the momentum representation,
\begin{equation}
{\cal D}_{\mu\nu}(x)=\int\frac{d^2 q_{\parallel} d^2
q_{\perp}}{(2\pi)^4}
e^{-iq_{\parallel}x_{\parallel}+iq_{\perp}x_{\perp}}
{\cal D}_{\mu\nu}\left(q_{\parallel},q_{\perp}\right).
\label{B7}
\end{equation}
After substituting this representation along with those in
Eqs.~(\ref{B4}) and (\ref{B5}) into the SD equation (\ref{B3})
and performing the straightforward, though tedious, integrations 
over $x_1^{\perp}$, $y_1^{\perp}$, $x_2^{\perp}$ and
$y_2^{\perp}$ we arrive at
\begin{eqnarray}
&&g\left(x_{\parallel}\right)=s\left(x_{\parallel}\right)
+4\pi \alpha \int \frac{d^4 q}{(2\pi)^4}
d^2 x_1^{\parallel} d^2 y_1^{\parallel}
\exp\left(-\frac{(q_{\perp}l)^2}{2}
-iq_{\parallel}(x_1^{\parallel}-y_1^{\parallel})\right)
s(x^{\parallel}-x_1^{\parallel})\gamma^{\mu}_{\parallel}
g(x_1^{\parallel}-y_1^{\parallel})\gamma^{\nu}_{\parallel}
g(y_1^{\parallel})
{\cal D}_{\mu\nu}\left(q_{\parallel},q_{\perp}\right)\nonumber\\
&&+(4\pi \alpha)^2 \int
\frac{d^4 q_1 d^4 q_2}{(2\pi)^8}
d^2 x_1^{\parallel} d^2 y_1^{\parallel}
d^2 x_2^{\parallel} d^2 y_2^{\parallel}
\exp\left(-\frac{(q_1^{\perp}l)^2}{2}-\frac{(q_2^{\perp}l)^2}{2}
+il^2[q_1^{\perp}\times q_2^{\perp}]
-iq_1^{\parallel}(x_1^{\parallel}-y_2^{\parallel})
-iq_2^{\parallel}(x_2^{\parallel}-y_1^{\parallel})
\right) \nonumber\\
&&\times
s(x^{\parallel}-x_1^{\parallel})\gamma^{\mu}_{\parallel}
g(x_1^{\parallel}-x_2^{\parallel})\gamma^{\sigma}_{\parallel}
g(x_2^{\parallel}-y_2^{\parallel})\gamma^{\nu}_{\parallel}
g(y_2^{\parallel}-y_1^{\parallel})\gamma^{\rho}_{\parallel}
g(y_1^{\parallel})
{\cal D}_{\mu\nu}\left(q_1^{\parallel},q_1^{\perp}\right)
{\cal D}_{\sigma\rho}\left(q_2^{\parallel},q_2^{\perp}\right),
\label{B8}
\end{eqnarray}
where $[q_1^\perp\times q_2^\perp]\equiv
\epsilon_{3ij}q_{1i}q_{2j}$.
Since  no dependence on $x_{\perp}$ has left, we conclude that
the form of $\tilde{G}(x)$ in Eq.~(\ref{B4}) is indeed consistent
with the structure of the SD equation.

Regarding this equation, it is necessary to emphasize that the
``perpendicular" components of the $\gamma$-matrices are absent
in it. Indeed, because of the identity  $O^{(-)}
\gamma_{\perp}^{\mu}O^{(-)}=0$, all those components are killed
by the projection operators coming from the fermion propagators.

Substituting now the photon propagator in the Feynman-like
(non-covariant) gauge (\ref{non-l}) into the SD equation, we see
that only the first term in Eq.~(\ref{non-l}), proportional to
$g^{\parallel}_{\mu\nu}$, leads to a nonvanishing contribution.
In other words, the photon propagator is effectively proportional
to $g^{\parallel}_{\mu\nu}$ (justifying the name of the gauge).

By switching to the momentum space, we obtain
\begin{eqnarray}
&&g^{-1}\left(p_{\parallel}\right)
=s^{-1}\left(p_{\parallel}\right)
-4\pi \alpha \int \frac{d^4 q}{(2\pi)^4}
\exp\left(-\frac{(q_{\perp}l)^2}{2}\right)
\gamma^{\mu}_{\parallel}
g(p^{\parallel}-q^{\parallel})\gamma^{\nu}_{\parallel}
{\cal D}_{\mu\nu}\left(q_{\parallel},q_{\perp}\right)\nonumber\\
&&-(4\pi \alpha)^2 \int
\frac{d^4 q_1 d^4 q_2}{(2\pi)^8}
\exp\left(-\frac{(q_1^{\perp}l)^2}{2}-\frac{(q_2^{\perp}l)^2}{2}
+il^2[q_1^{\perp}\times q_2^{\perp}] \right) \nonumber\\
&&\times
\gamma^{\mu}_{\parallel}
g(p^{\parallel}-q_1^{\parallel})\gamma^{\sigma}_{\parallel}
g(p^{\parallel}-q_1^{\parallel}-q_2^{\parallel})
\gamma^{\nu}_{\parallel}
g(p^{\parallel}-q_2^{\parallel})\gamma^{\rho}_{\parallel}
{\cal D}_{\mu\nu}\left(q_1^{\parallel},q_1^{\perp}\right) 
{\cal D}_{\sigma\rho}\left(q_2^{\parallel},q_2^{\perp}\right).
\label{B11}
\end{eqnarray}
The general solution to this equation is given by the ansatz,
\begin{equation}
g\left(p_{\parallel}\right)=
\frac{A_{p}\hat{p}_{\parallel}+B_{p}}
{A_{p}^2 p_{\parallel}^2-B_{p}^2},
\label{B12}
\end{equation}
where $A_{p}=A(p_{\parallel}^2)$ and $B_{p}=B(p_{\parallel}^2)$.
Making use of this as well as of the explicit form of the photon
propagator, the previous equation splits into the system of two
coupled equations,
\begin{eqnarray}
A_{p}&=&1+\left(\frac{\alpha}{2\pi^3}\right)^2 \int
\frac{d^2 q_1^{\parallel} d^2 q_2^{\parallel}
B_{p-q_1}A_{p-q_1-q_2}B_{p-q_2} }
{\left(A_{p-q_1}^2 (p^{\parallel}-q_1^{\parallel})^2
-B_{p-q_1}^2\right)
\left(A_{p-q_1-q_2}^2 (p^{\parallel}-q_1^{\parallel}
-q_2^{\parallel})^2
-B_{p-q_1-q_2}^2\right)
\left(A_{p-q_2}^2 (p^{\parallel}-q_2^{\parallel})^2
-B_{p-q_2}^2\right)}
\nonumber\\
&&\times  \int \frac{d^2 q_1^{\perp} d^2 q_2^{\perp}
\exp\left(-(q_1^{\perp}l)^2/2-(q_2^{\perp}l)^2/2
+il^2[q_1^{\perp}\times q_2^{\perp}] \right) }
{\left[q_1^2+(q_1^{\parallel})^2 \Pi
\left((q_1^{\perp})^2,(q_1^{\parallel})^2\right)\right]
\left[q_2^2+(q_2^{\parallel})^2 \Pi
\left((q_2^{\perp})^2,(q_2^{\parallel})^2\right)\right]},
\label{B13}\\
B_{p}&=&-i\frac{\alpha}{2\pi^3} \int 
\frac{d^2 q_{\parallel}B_{p-q}}
{\left(p_{\parallel}-q_{\parallel}\right)^2-B_{p-q}^2}
\int \frac{d^2 q_{\perp} \exp\left(-(q_{\perp}l)^2/2\right) }
{q^2+q_{\parallel}^2\Pi\left(q_{\perp}^2,q_{\parallel}^2\right)}
\nonumber\\
&&-\left(\frac{\alpha}{2\pi^3}\right)^2 \int
\frac{d^2 q_1^{\parallel} d^2 q_2^{\parallel}
(p_{\parallel}^2+q_1^{\parallel}\cdot q_2^{\parallel})
A_{p-q_1}B_{p-q_1-q_2} A_{p-q_2}  } {\left(A_{p-q_1}^2 
(p^{\parallel}-q_1^{\parallel})^2-B_{p-q_1}^2\right)
\left(A_{p-q_1-q_2}^2 (p^{\parallel}-q_1^{\parallel}
-q_2^{\parallel})^2 -B_{p-q_1-q_2}^2\right)
\left(A_{p-q_2}^2 (p^{\parallel}-q_2^{\parallel})^2 
-B_{p-q_2}^2 \right)} \nonumber\\
&&\times  \int \frac{d^2 q_1^{\perp} d^2 q_2^{\perp}
\exp\left(-(q_1^{\perp}l)^2/2-(q_2^{\perp}l)^2/2
+il^2[q_1^{\perp}\times q_2^{\perp}] \right) }
{\left[q_1^2+(q_1^{\parallel})^2 \Pi
\left((q_1^{\perp})^2,(q_1^{\parallel})^2\right)\right]
\left[q_2^2+(q_2^{\parallel})^2 \Pi
\left((q_2^{\perp})^2,(q_2^{\parallel})^2\right)\right]}.
\label{B14}
\end{eqnarray}
This concludes our derivation of the two-loop SD equations for
$A_{p}$ and $B_{p}$ which define the fermion propagator through
Eq.~(\ref{B12}).

Let us now show that the improved rainbow (one-loop)
approximation is reliable in gauge (\ref{non-l}). In that
approximation, the SD equations (\ref{B13}) and (\ref{B14}) are
\begin{eqnarray}
&&A_p=1,\label{oneloopA}\\
&&B_{p}=-i\frac{\alpha}{2\pi^3} \int
\frac{d^2 q_{\parallel}B_{p-q}}
{\left(p_{\parallel}-q_{\parallel}\right)^2-B_{p-q}^2}
\int \frac{d^2 q_{\perp} \exp\left(-(q_{\perp}l)^2/2\right) }
{q^2+q_{\parallel}^2\Pi\left(q_{\perp}^2,q_{\parallel}^2\right)}.
\label{oneloopB}
\end{eqnarray}
Equation (\ref{oneloopB}) was solved both numerically and
analytically (see the next section and Appendix~\ref{secB}). In
particular, it was shown that the approximation with
$B(p_\parallel^2)=m_{\rm dyn}$ for $p_\parallel^2<2|eB|$ and
$B(p_\parallel^2)$ rapidly decreasing for $p_\parallel^2>2|eB|$
is a very good one (see Fig.~\ref{fig-b} in the next Section).
Moreover, as it is shown in Sec.\ref{secV} and
Appendix~\ref{secB}, in the improved rainbow approximation, it is
sufficient to use the constant photon mass approximation for the
polarization function:  $q_\parallel^2 \Pi(q_\perp^2,
q_\parallel^2) \simeq -M_\gamma^2 \equiv -2\bar\alpha|eB|/\pi$
[see Eq.~(\ref{Pi-UV})]. Then, equation (\ref{B14}) at
$p_\parallel=0$ reduces to
\begin{equation}
1\simeq\frac{\alpha}{4\pi}\ln\left(\frac{1}{(M_{\gamma}l)^2}
\right) \ln\left(\frac{M_{\gamma}^2}{m_{dyn}^2(m_{dyn}l)^2}
\right)
+\mbox{Const}\frac{\alpha^2}{8\pi^2}
\ln^2\left(\frac{1}{(M_{\gamma}l)^2}\right)
\ln\left(\frac{M_{\gamma}^2}{m_{dyn}^2(m_{dyn}l)^2}\right),
\label{B15} \end{equation} where the two terms in the right hand
side come out as the estimates of the following two integrals,
\begin{eqnarray}
I_1&=&\frac{\alpha}{2\pi}
\int\limits_{0}^{\infty}\frac{dx dy \exp(-x(m_{dyn}l)^2/2)}
{(y+1)(x+y+(M_{\gamma}/m_{dyn})^2)},\\
I_2&=&\frac{\alpha^2}{8\pi^2}\int\limits_{0}^{\infty}
\frac{dx_1 dy_1 dx_2 dy_2
\exp\left(-x_1(m_{dyn}l)^2/2-x_2(m_{dyn}l)^2/2\right)
J_{0}\left((m_{dyn}l)^2\sqrt{x_1 x_2}\right)}
{(y_1+1)(y_2+1)\left(x_1+y_1+(M_{\gamma}/m_{dyn})^2\right)
\left(x_2+y_2+(M_{\gamma}/m_{dyn})^2\right)}\nonumber \\
&&\times
\left(\frac{y_1+y_2+1}{\sqrt{(y_1-y_2)^2+2(y_1+y_2)+1}}-1\right).
\end{eqnarray}
By assuming that the first term on the right hand side in
Eq.~(\ref{B15}) is of order one, we see that the second term,
corresponding to the two-loop correction, is indeed suppressed:
it is of order $\alpha\log{1/\alpha}$. Note that we should have
expected this result taking into account the calculations in
Appendix~\ref{secA}. As is shown there, the higher order
corrections to the vertex are suppressed in the Feynman-like
gauge, and the two-loop contribution in the SD equation occurs as
a result of the one-loop insertion in the vertex (see
Fig.~\ref{fig-SD}).

If we  repeat the same analysis in the case of the covariant
gauge (\ref{cov-gauge}), we  end up with the following estimate,
\begin{eqnarray}
1&\simeq & \frac{\alpha}{4\pi}
\ln^2\left(\frac{1}{(m_{dyn}l)^2}\right)
+\mbox{Const}\alpha^2
\ln^4\left(\frac{1}{(m_{dyn}l)^2}\right),
\label{B16}
\end{eqnarray}
where the main contribution to both terms, as is easy to check,
comes from the $q^{\mu}_{\parallel} q^{\mu}_{\parallel}
/q^2q^2_{\parallel}$ component in the photon propagator. In
contrast to what we had in the Feynman-like gauge, after assuming
that the first term in Eq.~(\ref{B16}) is of order of one, we see
that the second term would be also of order one. In other words,
there is no suppression of higher order terms in the covariant
gauge. Similarly, there is no suppression in all other gauges in
which the $q^{\mu}_{\parallel} q^{\mu}_{\parallel}
/q^2q^2_{\parallel}$ component is non-zero. Therefore, in
agreement with the general arguments of the previous section, the
Feynman-like gauge (\ref{non-l}) is special: in this gauge there
exists a consistent truncation of the SD equations.

Since the dynamical mass is a gauge invariant quantity, in other
gauges one needs to sum up an infinite set of diagrams to recover
the same result for it. Obviously it is rather difficult to
classify all the relevant diagrams in those gauges, and so the
existence of the Feynman-like gauge (\ref{non-l}) is the key
point that allows to treat the present problem reliably.

\section{The analysis of the SD equations}
\label{secV}

As was shown in the previous section, in the gauge (\ref{non-l}),
the SD equations (\ref{G{-1}}), (\ref{SD-fer}), (\ref{SD-pho})
and (\ref{Pi_munu}) with the bare vertex (\ref{bare-ver}) are
reliable. They form a closed system of integral equations. In
Euclidean space Eq.~(\ref{oneloopB}) has the form
\begin{eqnarray}
B(p_{\parallel}^2)=\frac{\alpha}{2\pi^2}
\int\frac{d^2 q_{\parallel} B\left((p_{\parallel}
-q_{\parallel})^2\right)} {(p_{\parallel}-q_{\parallel})^2
+B^2\left((p_{\parallel} -q_{\parallel})^2\right)}
\int\limits_{0}^{\infty}\frac{dx \exp(-xl^2/2)}
{x+q_{\parallel}^2+q_{\parallel}^2\Pi_{E}(x,q_{\parallel}^2)},
\label{SD-B}
\end{eqnarray}
where the polarization function
$\Pi_{E}(q_{\perp}^2,q_{\parallel}^2)$ is defined from
Eq.~(\ref{Pi_munu}) with the bare vertex (\ref{bare-ver}).

Equation (\ref{SD-B}) was solved by using both analytical and
numerical methods. The analytical approach is considered in
Appendix~\ref{secB}. Here we will describe the numerical
solution.

\begin{figure}
\epsfbox[0 500 350 800]{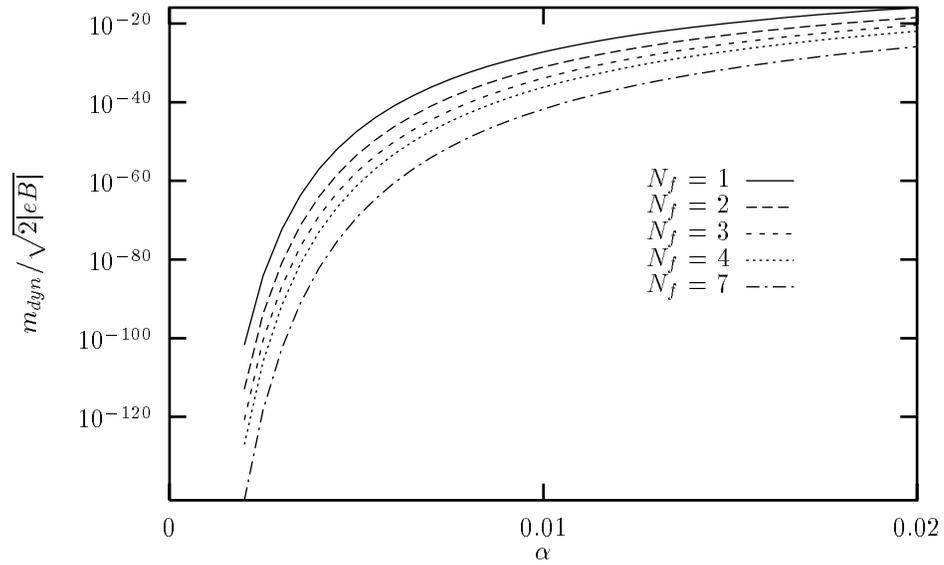}
\caption{Plot of $m_{dyn}$ as a function  of $\alpha$
for several values of $N_f$.}
\label{fig-m}
\end{figure}

\begin{figure}
\epsfbox[0 500 350 800]{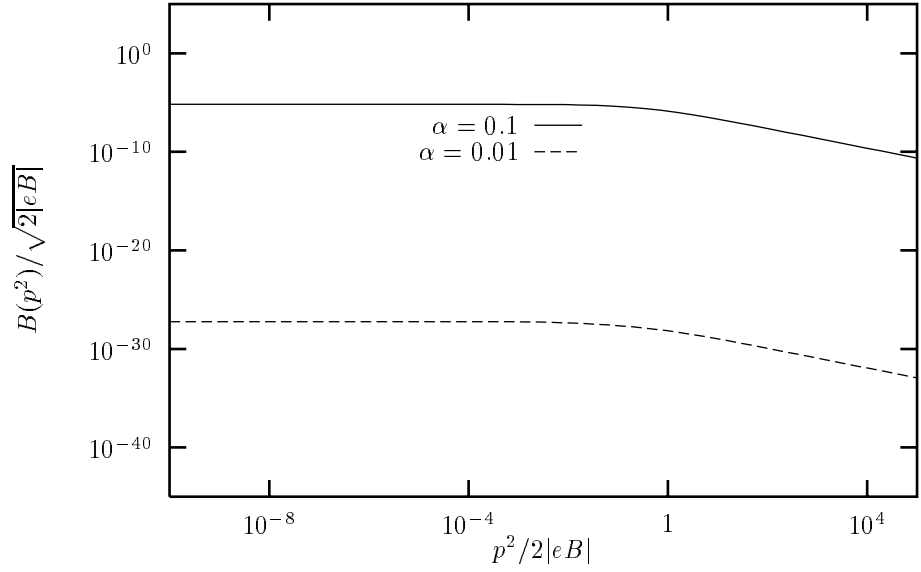}
\caption{Plot of the mass function $B(p^2)$ as a function  of $p^2$
for $N_f=1$ and two values of $\alpha$.}
\label{fig-b}
\end{figure}

\begin{figure}
\epsfbox[0 500 350 780]{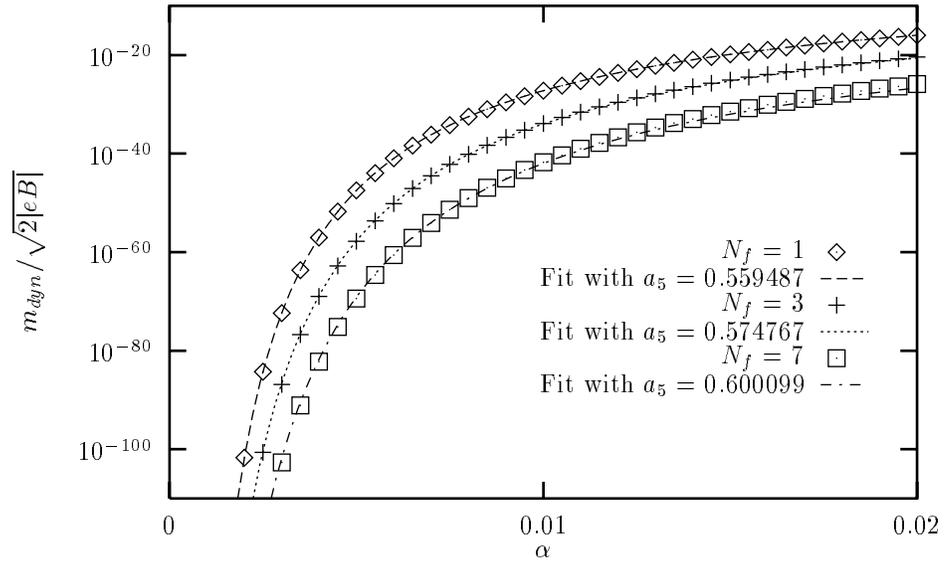}
\caption{Plot of the fit function and corresponding datapoints for
several values of $N_f$ when the only free parameter was $a_5$. }
\label{fig-fit}
\end{figure}

The results of the numerical analysis of Eq.~(\ref{SD-B}) are
shown in Figs.~\ref{fig-m}, \ref{fig-b} and \ref{fig-fit}. The
crucial points in the analysis are the following:

a) The polarization function $\Pi_{E}(q_{\perp}^2,
q_{\parallel}^2)$ is in principle a complicated functional of the
fermion mass function $B(p_{\parallel}^2)$. However, as it is
shown in  Appendix~\ref{secB}, the leading singularity,
$1/\alpha\ln(\alpha)$, in $\ln(m_{dyn}^2)$ in Eq.~(\ref{m}) is
induced in the kinematic region with $m_{dyn}^2 \ll
|q_{\parallel}^2| \ll |eB|$ and $m_{dyn}^2 \ll M_{\gamma}^2 \alt
q_{\perp}^2 \ll |eB|$. In that region, the fermions can be
treated as massless, and therefore the polarization function is
$\Pi_{E}(q_{\perp}^2,q_{\parallel}^2) \simeq 2\bar{\alpha} |eB|
/\pi q_{\parallel}^2 =M_{\gamma}^2/q_{\parallel}^2$ [see
Eqs.~(\ref{Pi-UV}) and (\ref{M_gamma})]. In other words, in this
approximation, the photon propagator is a propagator of a free
massive boson with $M_{\gamma}^2=2\bar{\alpha}|eB|/\pi$.

b) In the numerical solution, the following ansatz for $\ln
(m_{dyn})$ was used,
\begin{equation}
\ln\frac{m_{dyn}}{\sqrt{2|eB|}}=\ln a_0 +{a_1\over3}\ln
\frac{N\alpha}{\pi}-\frac{a_2}{(\frac{\alpha}{\pi})^{a_3}
\ln^{a_4} \frac{a_5\pi}{N\alpha.}}.
\end{equation}
For small $\alpha$ ($0.001\leq\alpha\leq 0.1$) and different $N$
($1\leq N\leq7$) the best fit was found with
$a_0=a_1=a_2=a_3=a_4=1$ and $a_5\simeq 0.58\pm 0.02$ (see
Fig.~\ref{fig-fit}). This fit corresponds to the expression
\begin{equation}
m_{dyn} =\tilde C \sqrt{|eB|} F(\alpha)\exp\left[-\frac{\pi}
{\alpha\ln\left(C_1/N\alpha\right)}\right], \label{dynmass}
\end{equation}
where $F(\alpha)\simeq(N\alpha)^{1/3}$, $C_1\simeq 1.82\pm 0.06$
and $\tilde C=\sqrt2$.

c) The numerical solution shows that the function
$B(p_{\parallel}^2)$ is essentially constant for
$p_{\parallel}^2 \ll |eB|$, $B(p_{\parallel}^2)=m_{dyn}$, and
rapidly decreases for $p_{\parallel}^2\gg |eB|$, see
Fig.~\ref{fig-b}. Therefore this approximation is
self-consistent: the Ward-Takahashi identity for the vertex is
satisfied in the relevant kinematic region of momenta (see
Appendix~\ref{secC}), and the pole of the fermion
propagator appears at $p_{\parallel}^2 =m_{dyn}^2$.

It is instructive to clarify the origin of the difference between
the dynamical mass $m_{dyn}$ (\ref{m_dyn}), calculated in the
rainbow approximation, and the expression for $m_{dyn}$
(\ref{dynmass}). The integral equation (\ref{SD-B}) for
$B(p_\parallel^2)$ can be rewritten in the form,
\begin{equation}
B(p^2)=\frac{\alpha}{2\pi^2}
\int\frac{d^2q B(q^2)} {q^2+B^2(q^2)}
\int\limits_{0}^{\infty}\frac{dx \exp(-xl^2/2)}
{x+({\bf q}-{\bf p})^2+M_\gamma^2},
\label{inteq:B}
\end{equation}
with $M_\gamma^2=0$ and $M_\gamma^2=2\bar\alpha|eB|/\pi$ in the
rainbow approximation and in the improved rainbow approximation,
in gauge (\ref{non-l}), respectively (henceforth we will omit the
symbol $\parallel$; the two-dimensional vector 
${\bf q}=(q_4,q_3), q_4=-iq_0$).

The numerical analysis of this integral equation shows that the
so called linearized approximation, with $B^2(q^2)\rightarrow
m_{\rm dyn}^2=B^2(0)$ in the denominator of Eq.~(\ref{inteq:B}),
is an excellent approximation. Then we get
\begin{equation}
B(p^2)=\frac{\alpha}{2\pi^2}
\int\frac{d^2q B(q^2)} {q^2+m_{dyn}^2}
\int\limits_{0}^{\infty}\frac{dx \exp(-xl^2/2)}
{x+({\bf q}-{\bf p})^2+M_\gamma^2}.
\label{inteqlinear}
\end{equation}
This equation is equivalent to a two-dimensional
Schr\"odinger-like differential equation. Indeed, introducing
the function 
\begin{eqnarray}
\Psi({\bf r})=\int\frac{d^2q}{(2\pi)^2}
\frac{B(q^2)}{q^2+m_{dyn}^2} e^{i{\bf qr}},
\end{eqnarray}
we get the equation
\begin{equation}
(-\Delta+m^2_{dyn}+V({\bf r}))\Psi({\bf r})=0,
\label{eq:schre}
\end{equation}
where $\Delta =\partial^2/\partial x^2_3+\partial^2/\partial
x^2_4$ and the potential $V({\bf r})$ is
\begin{eqnarray}
V({\bf r})&=&-\frac{\alpha}{2\pi^2}\int d^2p e^{i{\bf pr}}
\int\limits^{\infty}_{0} \frac{dx \exp(-x/2)}
{l^2p^2+x+l^2M_\gamma^2}  \nonumber\\
&=&-\frac{\alpha}{\pi l^2}\int\limits^{\infty}_{0} dx e^{-x/2}
K_{0}\left(\frac{r}{l}\sqrt{x+l^2M_\gamma^2}\right).
\end{eqnarray}
At $M_\gamma^2=0$ (the rainbow approximation), the last integral
is expressed through the integral exponential function
$Ei(x)=-\int_{-x}^{\infty}dt \exp(-t)/t$,
\begin{eqnarray}
V_0({\bf r})\equiv V({\bf r})\Bigg|_{M_\gamma=0}
=\frac{\alpha}{\pi l^2} \exp\left(\frac{r^2}{2l^2}\right)
Ei\left(-\frac{r^2}{2l^2}\right).
\label{eq:pot}
\end{eqnarray}
Using the asymptotic relations for $Ei(x)$ \cite{Ryz}, we get
\begin{mathletters}
\begin{eqnarray}
V_0({\bf r})&\simeq &-\frac{2\alpha}{\pi}\frac{1}{r^2},
\qquad r\gg l;\label{60a}\\
V_0({\bf r})&\simeq &-\frac{\alpha}{\pi l^2}
\log\frac{2l^2}{r^2}, \qquad r\ll l.
\label{60b}
\end{eqnarray}
\label{60}
\end{mathletters}
Equation (\ref{60a}) implies that the potential is long-range in
the rainbow approximation.

Using now the asymptotic relations for $K_0(z)$ \cite{Ryz}, we
get the following asymptotes for $V({\bf r})$ at
$M_\gamma^2\neq0$, 
\begin{mathletters}
\begin{eqnarray}
V({\bf r})&\simeq&-\sqrt{\frac{2}{\pi}}
\alpha\frac{M_\gamma^{1/2}}
{r^{3/2}} e^{-M_\gamma r},\quad r\gg\frac{1}{M_\gamma}\gg
l,\label{61a}\\
V({\bf r})&\simeq&-\frac{\alpha}{\pi
l^2}\log\frac{2l^2}{r^2},\quad r\ll l.
\label{61b}
\end{eqnarray}
\label{61}
\end{mathletters}
Therefore, while the short-distance behavior of the potential is
independent of $M_\gamma$ (compare Eqs.~(\ref{60b}) and
(\ref{61b})), its long-distance behavior in the rainbow
approximation and in the improved rainbow approximation
 is essentially different: while the former potential is
long-range [see Eq.~(\ref{60a})], the latter is short-range [see
Eq.~(\ref{61a})]. This point yields the physical origin of the
difference of expressions (\ref{m_dyn}) and (\ref{dynmass}) for
$m_{dyn}$ in those two approximations.

It is also instructive to discuss this point using results proved
in the literature for the two-dimensional ($d=2$) Schr\"odinger
equation \cite{Simon}. The form of equation (\ref{eq:schre})
implies that $-m_{dyn}^2$ plays the role of energy $E$ in the
Schr\"odinger equation. The results of Ref.\cite{Simon} ensure
that a) there is at least one bound state for any attractive
potential for $d=2$, {\em i.e.}, there is at least one solution
with $m_{\rm dyn}^2=-E>0$; b) for short-range potentials of the
form $V({\bf r})=\alpha\tilde{V}({\bf r})$, where 
$\tilde{V}({\bf r})$ is independent of $\alpha$, the energy
$E=-m_{dyn}^2$ of the lowest bound state is $-E(\alpha)
=m_{\rm dyn}^2(\alpha) \sim\exp(-1/{a\alpha})$, with $a>0$, as
$\alpha\to0$. If the mass $M_\gamma$ were independent of
$\alpha$, the potential (\ref{61a}) would satisfy the
constraints in item b) above. However, because
$M_\gamma^2=2N\alpha|eB|/\pi\to0$ as $\alpha\to0$, we get an
additional logarithmic factor in the power of the exponent in
equation (\ref{dynmass}). On the other hand, since the potential
$V({\bf r})$ is long-range in the rainbow approximation [see
Eq.~(\ref{60a})], it clearly does not satisfy the constraints in
item b) above, and $m_{dyn}^2 \sim \exp(-1/{a\sqrt{\alpha}})$ in
that approximation.\footnote{As follows from the analysis in
Appendix~\ref{secB}, $m_{dyn}$ would be indeed $m_{dyn}\sim
\sqrt{|eB|}\exp(-1/{a\alpha})$, with $a\sim\log{|eB|/
M_\gamma^2}$, in the case of $M_\gamma^2$ independent of
$\alpha$. Notice that the transition from $M_\gamma^2
=2N\alpha|eB|/\pi$ (the improved rainbow approximation) to
$M_\gamma^2=0$ (the rainbow approximation) corresponds to
changing $a\sim\log{|eB|/M_\gamma^2} \sim\log{1/N\alpha}$ to
$a \sim\log{|eB|/m_{dyn}^2}$. Then, expression (\ref{dynmass})
for $m_{dyn}$ transforms into expression (\ref{m_dyn}).}

\section{Conclusion}
\label{secVI}

The magnetic catalysis of chiral symmetry breaking in QED is
a phenomenon with rather rich and sophisticated dynamics. It
yields a (first, to the best of our knowledge) example in which
there exists a consistent truncation of the Schwinger-Dyson
equations in the problem of dynamical symmetry breaking in a
(3+1)-dimensional gauge theory without fundamental scalar fields.

It is instructive to compare this problem with dynamical symmetry
breaking in (3+1)-dimensional gauge theories without external
fields (for a review, see Refs. \cite{22,FJ}). In the case of
non-Abelian gauge theories, such as QCD, dynamical chiral
symmetry breaking is generated in the infrared region, where
the effective coupling constant is strong. This prevents to
elaborate a consistent truncation of the Schwinger-Dyson
equations in those theories. In Abelian gauge theories, on the
other hand, a solution with a non-zero fermion mass exists in
ladder approximation for any value of the coupling constant
\cite{JJ}, {\em if} there is no ultraviolet cutoff in the
Schwinger-Dyson equations (at finite cutoff, in the ladder
approximation, dynamical chiral symmetry breaking takes place
only if the coupling constant is large enough \cite{22}).
This fact implies that the ultraviolet region is responsible for
chiral symmetry breaking in that case. Since the running coupling
is strong in the ultraviolet region in Abelian gauge theories,
this again prevents to elaborate a consistent truncation of the
Schwinger-Dyson equations in those theories either. In contrast,
the dynamics of chiral symmetry breaking in QED in a magnetic
field is long-range, and the QED coupling constant is small in
infrared. As a result, a consistent truncation of the SD
equations exists in the present problem in the gauge
(\ref{non-l}).

The crucial point in the present analysis is the dimensional
reduction in the dynamics of the fermion pairing. However, there
is an essential difference between QED and the Nambu-Jona-Lasinio
model in a magnetic field: in QED, there is an additional neutral
field, the photon field. As a result, the dynamics in QED in a
magnetic field is much more sophisticated than that in the NJL
model. In particular, the photon propagator includes the
four-dimensional $q^2=q_{\parallel}^2-q_\perp^2$ and is not
reduced to the two-dimensional form. However, the tensor and the
spinor structure in this model in the infrared region is exactly
the same as in the Schwinger model. This point is crucial for
making this problem essentially soluble.

What is the chiral symmetry, $SU_{L}(N) \times SU_{R}(N) \times
U_{V}(1)$ or $U_L(N)\times U_R(N)$, in this problem? It is known
that in massless QED, without external fields, the chiral
symmetry is $U_L(N)\times U_R(N)$: though the singlet axial
current $j_5^\mu$ is not conserved, the corresponding charge
$Q_5$ is conserved. The latter is connected with the absence of
instanton-like configurations in QED. In the presence of an
external magnetic field, the situation however might be
different. Indeed, the dynamics in QED in a magnetic field is
intimately connected with the Schwinger model where the $U_A(1)$
symmetry is explicitly broken. We believe that this issue
deserves further study.

An important ingredient of the dynamics in QED in a magnetic
field is the pseudo-Higgs effect. It is not a genuine Higgs
effect since there is no complete screening of the electric
charge: Eq.~(\ref{Pi-IR}) implies that in the deep infrared
region with $|q_{\parallel}^2|\ll m_{dyn}^2$, there are ordinary
Coulomb-like forces. Still, the pseudo-Higgs effect is manifested
in creating a resonance in the photon channel, with
$M_\gamma^2=2\bar\alpha|eB|/\pi\gg m_{dyn}^2$. This resonance
provides the dominating forces leading to chiral symmetry
breaking.

Thus, in this problem, the region primarily responsible for
chiral symmetry breaking is the region of intermediate momenta:
$m_{dyn}^2\ll |q_{\parallel}^2|\ll |eB|$ and $m_{dyn}^2\ll
M_\gamma^2 \alt |q_{\perp}^2|\ll |eB|$. This point is noticeable
as an example for a possibility discussed for the QCD dynamics
\cite{22}: the dynamics of spontaneous chiral symmetry breaking
might be provided by forces essentially independent of the
dynamics of confinement (the infrared dynamics).

Another noticeable point is the existence of a special gauge in
which the description of the nonperturbative dynamics is
essentially simplified. Recall that the conventional viewpoint
now is that there is a particular gauge in QCD (the maximal
Abelian gauge) which is the most appropriate one for the
description of confinement  and chiral symmetry breaking
\cite{thooft}.

At last, there are arguments in support of a dimensional
reduction in the dynamics of chiral symmetry breaking in QCD
\cite{VafaWitten}. The present model yields an example of such a
mechanism.

We hope that the dynamics of the magnetic catalysis of chiral
symmetry breaking in QED will provide insight into the
non-perturbative dynamics of more complicated theories, such as
quantum chromodynamics.

\begin{acknowledgments}
We are grateful to Anthony Hams for his generous help in
numerical solving the Schwinger-Dyson equations. We thank
H.~Minakata, A.~Smilga, L.C.R.~Wijewardhana and K.~Yamawaki for
useful discussions. The work of V.P.G. was supported by Swiss
National Science Foundation Grant No. CEEC/NIS/96-98/7 IP 051219
and Foundation of Fundamental Researches of Ministry of Science of
Ukraine under Grant No.2.5.1/003. The work of I.A.S. was
supported by U.S. Department of Energy  Grant \#DE-FG02-84ER40153.
\end{acknowledgments}

\appendix
\section{Loop corrections to the vertex}
\label{secA}

\begin{figure}
\epsfbox{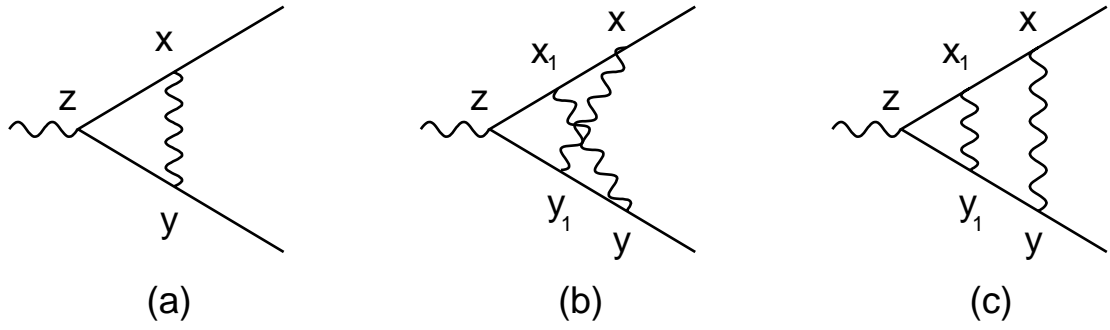}
\caption{The diagrammatic representation of the one- and
two-loop corrections to the vertex.}
\label{fig-ver}
\end{figure}

In this Appendix we will calculate the one and two-loop
corrections to the vertex in the gauge (\ref{non-l}) (see
Fig.\ref{fig-ver}). In agreement with the general arguments in
Sec.~\ref{secIII}, it will be shown that they are small: of order
$\alpha$ and $\alpha^2$ in one-loop and two-loop approximations,
respectively. On the other hand, we will show that in covariant
gauges (\ref{cov-gauge}), there is a large, $O(1)$, correction to
the vertex already in the one-loop approximation.

As was pointed out in Sec.~\ref{secIII}, the loop expansion we
use is based on the effective action of Cornwall, Jackiw, and
Tomboulis. In this expansion, the full fermion and photon
propagators are used and only two-particle irreducible graphs
have to be taken into account \cite{21,22}. Our aim is to show
that in the gauge (\ref{non-l}), the improved rainbow
approximation is reliable, {\em i.e.}, the loop corrections to
the vertex, with the photon and fermion propagators calculated in
that approximation, are small. In this connection, notice that in
the improved rainbow approximation the function
$A(p_{\parallel}^2)$ in the fermion propagator equals one [see
Eq.~(\ref{oneloopA})] and $B(p_{\parallel}^2)\simeq m_{dyn}$ for
all $p_{\parallel}^2 \alt 2|eB|$ (see Fig.~\ref{fig-b}).
Therefore one can use the bare propagator (\ref{tildeS}) with
$m=m_{dyn}$ in the calculations of the loop corrections to the
vertex: indeed, as was indicated in Sec.~\ref{secIII}, the
perturbative expansion in $\alpha$ potentially might be destroyed
by mass singularities coming from infrared region, where
$B(p_{\parallel}^2)\simeq m_{dyn}$.

After getting rid of the Schwinger phase factors according to the
prescription in Eq.~(\ref{14}), we get the following expression
for the one-loop correction to the vertex function,
\begin{equation}
\tilde{\Gamma}^{(1)\mu}(x-z,y-z)=(ie)^2
\exp\left(i(x-y)^{\lambda}A_{\lambda}(z-y)\right)
\gamma^{\lambda}\tilde{S}^{(L)}(x-z)\gamma^{\mu}
\tilde{S}^{(L)}(z-y)\gamma^{\nu}{\cal D}_{\lambda\nu}(x-y)
\label{A3}
\end{equation}
(see Fig.~\ref{fig-ver}a). By performing the Fourier transform in
both $x-z$ and $y-z$, we arrive at
\begin{eqnarray}
\tilde{\Gamma}^{(1)\mu}(p,k)&=& (ie)^2 \int d^4 x  d^4 y
\frac{d^4 k_1 d^4 k_2 d^4 q}{(2\pi)^{12}}
 e^{ixp-iyk-ixA(y)-ik_1 x+ik_2 y +iq(x-y)}
\gamma^{\lambda}\tilde{S}^{(L)}(k_1)\gamma^{\mu}
\tilde{S}^{(L)}(k_2)\gamma^{\nu}{\cal D}_{\lambda\nu}(q)
\nonumber\\
&=&(ie)^2\int d^4 x
\frac{d^4 k_1 d^4 q}{(2\pi)^8} e^{-ix k_1}
\gamma^{\lambda}\tilde{S}^{(L)}(k_1+q+p)\gamma^{\mu}
\tilde{S}^{(L)}(q+k-\vec{A}(x))\gamma^{\nu}
{\cal D}_{\lambda\nu}(q)
\label{A4}
\end{eqnarray}
where we first integrated over $y$ and $k_2$, and then
shifted $k_1$ by $p+q$.

Substituting the fermion propagator as in Eq.~(\ref{tildeS}),
but with $m=m_{dyn}$, into the last expression,
we obtain
\begin{eqnarray}
\tilde{\Gamma}^{(1)\mu}(p,k)&=&\frac{\alpha}{2\pi^3} \int
\frac{d^2 q_{\parallel} d^2 q_{\perp}
\exp\left[-(q_{\perp}+p_{\perp})^2l^2 /2
-(q_{\perp}+k_{\perp})^2l^2 /2
+il^2 [q_{\perp}\times (p_{\perp}-k_{\perp})]
+il^2 [p_{\perp}\times k_{\perp}]\right]}
{\left[(q_{\parallel}+p_{\parallel})^2-m_{dyn}^2\right]
\left[(q_{\parallel}+k_{\parallel})^2-m_{dyn}^2\right]}
\nonumber\\
&\times&\gamma^{\lambda}
(\hat{q}_{\parallel}+\hat{p}_{\parallel}+m_{dyn}) O^{(-)}
\gamma_{\parallel}^{\mu}
(\hat{q}_{\parallel}+\hat{k}_{\parallel}+m_{dyn}) O^{(-)}
\gamma^{\nu} {\cal D}_{\lambda\nu}(q_{\parallel},q_{\perp}),
\label{A5}
\end{eqnarray}
where $[q^\perp\times p^\perp]\equiv \epsilon_{3ij}q_{i}p_{j}$
(with $i,j=1,2$). As is clear from this expression, the
dependence on the perpendicular momenta is weak in infrared,
$0<|p_{\perp}|, |k_{\perp}|\alt 1/l$, and exponentially
suppressed in ultraviolet. For our purposes, it is sufficient
to use $|p_{\perp}|, |k_{\perp}|=0$. Regarding the
parallel components, we keep them non-zero so far.

After substituting the photon propagator (\ref{non-l}) into
Eq.~(\ref{A5}), we notice that the result contains two kinds of
terms, namely, terms proportional to $O^{(+)} =(1+i\gamma^1
\gamma^2 {\rm sign}(eB))/{2}$ and terms proportional to
$O^{(-)}=(1-i\gamma^1\gamma^2{\rm sign})(eB)/{2}$. As we discuss
in Sec.~\ref{secIV}, the former are completely irrelevant for the
SD equation and we drop them. The latter are relevant and has to
be carefully analyzed. As is easy to check, they come exclusively
from the $g_{\mu\nu}^{\parallel}$ term in the photon propagator,
and the explicit expression (in Euclidean space) reads
\begin{eqnarray}
\tilde{\Gamma}^{(1)\mu}(p_{\parallel},k_{\parallel})&\simeq&
\frac{\alpha m_{dyn}}{\pi^3} \int
\frac{d^2 q_{\parallel} d^2 q_{\perp} e^{-(q_{\perp}l)^2}
\left[\gamma_{\parallel}^{\mu}
(\hat{q}_{\parallel}+\hat{p}_{\parallel})
+(\hat{q}_{\parallel}+\hat{k}_{\parallel})
\gamma_{\parallel}^{\mu}\right] O^{(-)}  }
{\left[(q_{\parallel}+p_{\parallel})^2+m_{dyn}^2\right]
\left[(q_{\parallel}+k_{\parallel})^2+m_{dyn}^2\right]
\left(q_{\parallel}^2+q_{\perp}^2+M_{\gamma}^2\right)}.
\label{A5a}
\end{eqnarray} 
In a standard way, we introduce the Feynman parameters and
perform the integration over $q_{\parallel}$. The result reads,
\begin{eqnarray}
&&\tilde{\Gamma}^{(1)\mu}(p_{\parallel},k_{\parallel})\simeq
\frac{\alpha}{\pi} m_{dyn} O^{(-)} \nonumber\\
&\times&\int\limits_{0}^{\infty} d z e^{-zl^2}
\int\limits_{0}^{1} d x \int\limits_{0}^{1-x} d y
\frac{ (1-y) \hat{k}_{\parallel} \gamma_{\parallel}^{\mu}
-y \gamma_{\parallel}^{\mu} \hat{k}_{\parallel}
-x \hat{p}_{\parallel} \gamma_{\parallel}^{\mu}
+(1-x) \gamma_{\parallel}^{\mu} \hat{p}_{\parallel}  }
{\left[x(p_{\parallel}^2+m_{dyn}^2)+y(k_{\parallel}^2+m_{dyn}^2)
-(xp_{\parallel}+yk_{\parallel})^2 +(1-x-y)(z+M_{\gamma}^2)
\right]^2}.
\label{A5b}
\end{eqnarray}
Note that this is zero for $k_{\parallel}=p_{\parallel}=0$,
and that the integral in the right hand side is finite for any
finite $m_{dyn}^2$. In the most interesting region of momenta,
$m^2\ll p_{\parallel}^2, k_{\parallel}^2\ll 1/l^2$, this integral
contains the logarithmic contribution, $\ln (m_{dyn}l)^2$.

While estimating the effect of the one-loop correction in
Eq.~(\ref{A5b}) to the SD equation in Sec.~\ref{secIV}, we
could check that the most important contribution comes from
the region of momenta (up to the exchange of $p_{\parallel}$
and $k_{\parallel}$) $m^2\ll p_{\parallel}^2 \ll 1/l^2$ and
$k_{\parallel}\simeq 0$. In this particular case, from
Eq.~(\ref{A5b}) we obtain
\begin{eqnarray}
&&\tilde{\Gamma}^{(1)\mu}(p_{\parallel},0)\simeq
\frac{\alpha}{\pi} m_{dyn} O^{(-)}
\int\limits_{0}^{\infty} \frac{ d z e^{-zl^2} }{z+M_{\gamma}^2}
\nonumber\\ &\times& \left( \frac{\gamma_{\parallel}^{\mu}
\hat{p}_{\parallel}}{p_{\parallel}^2} 
\ln\frac{(p_{\parallel}^2)^2}
{m_{dyn}^2(z+M_{\gamma}^2+p_{\parallel}^2)}
-\frac{\hat{p}_{\parallel}
\gamma_{\parallel}^{\mu} }{p_{\parallel}^2}
\ln\frac{p_{\parallel}^2(z+M_{\gamma}^2+p_{\parallel}^2)}
{m_{dyn}^2(z+M_{\gamma}^2)}
+\frac{\hat{p}_{\parallel} \gamma_{\parallel}^{\mu} }
{z+M_{\gamma}^2+p_{\parallel}^2}
\ln\frac{(z+M_{\gamma}^2+p_{\parallel}^2)^2}
{m_{dyn}^2(z+M_{\gamma}^2)} \right) \nonumber\\
&\simeq&  \frac{\alpha}{\pi} m_{dyn} O^{(-)}
\left( \frac{\gamma_{\parallel}^{\mu} 
\hat{p}_{\parallel}}{p_{\parallel}^2}
\ln\frac{p_{\parallel}^2}{m_{dyn}^2}
\ln\frac{1}{(M_{\gamma}l)^2} -\frac{\hat{p}_{\parallel} 
\gamma_{\parallel}^{\mu}}{p_{\parallel}^2}
\ln\frac{1}{(m_{dyn}l)^2}
\ln\frac{1}{(M_{\gamma}l)^2+(p_{\parallel}l)^2}
\right).
\label{A5c}
\end{eqnarray}
This, in its turn, results in a suppressed two-loop correction to
the SD equation.

Now, let us consider the correction to vertex function at
two-loop order. There are two diagrams in this order (see
Fig.~\ref{fig-ver}). Here we shall explicitly describe the
correction connected with the diagram with two crossed photon
lines (Fig.~\ref{fig-ver}b); the analysis of another correction,
connected with the diagram with two parallel photon lines
(Fig.~\ref{fig-ver}c) can be done similarly.

In coordinate space, the expression corresponding to the diagram
with two crossed photon lines reads
\begin{eqnarray}
\tilde{\Gamma}^{(2)\mu}(x,y)&=&(ie)^4 \int d^4 x_1 d^4 y_1
\exp\left[ix^{\lambda}A_{\lambda}(x_1)
+iy_1^{\lambda}A_{\lambda}(y)+iy^{\lambda}A_{\lambda}(x)\right]
\gamma^{\sigma}\tilde{S}^{(L)}(x-x_1)
\gamma^{\nu}\tilde{S}^{(L)}(x_1) \nonumber\\
&\times&\gamma^{\mu}\tilde{S}^{(L)}(-y_1)
\gamma^{\rho}\tilde{S}^{(L)}(y_1-y)
\gamma^{\lambda}{\cal D}_{\nu\lambda}(x_1-y)
{\cal D}_{\sigma\rho}(x-y_1).
\label{A7}
\end{eqnarray}
In the momentum space, this becomes
\begin{eqnarray}
\tilde{\Gamma}^{(2)\mu}(p,k)&=&
(ie)^4 \int d^4 x d^4 y d^4 x_1 d^4 y_1
\frac{d^4 k_1 d^4 k_2 d^4 k_3 d^4 k_4 d^4 q_1 d^4 q_2}
{(2\pi)^{24}} \nonumber\\&\times&
e^{ixp-iyk+ixA(x_1)+iy_1 A(y)-ixA(y)-ik_1 (x-x_1)-ik_2 x_1
+ik_3 y_1 -ik_4(y_1-y)+iq_1(x_1-y)+iq_2(x-y_1)} \nonumber\\
&\times& \gamma^{\sigma}\tilde{S}^{(L)}(k_1)
\gamma^{\nu}\tilde{S}^{(L)}(k_2)
\gamma^{\mu}\tilde{S}^{(L)}(k_3)
\gamma^{\rho}\tilde{S}^{(L)}(k_4) \gamma^{\lambda}
{\cal D}_{\nu\lambda}(q_1) {\cal D}_{\sigma\rho}(q_2).
\end{eqnarray}
Performing the straightforward integrations over $x_1$ and $y_1$,
and then over $k_i$ (by making use of the explicit form of the
fermion propagator), at the end we arrive at the following
expression for the two-loop vertex correction at $k=p=0$ (the
case $k,p\neq 0$ is not expected to give a very different
estimate, but it is much harder to work with),
\begin{eqnarray}
\tilde{\Gamma}^{(2)\mu}(0,0)&=&\frac{\alpha^2}{\pi^6}
\int \frac{d^2 q_{1}^{\parallel} d^2 q_{2}^{\parallel}
d^2 q_{1}^{\perp} d^2 q_{2}^{\perp}
\exp\left[-\frac{3}{2}(q_{1}^{\perp}l)^2
-\frac{3}{2}(q_{2}^{\perp}l)^2
+2i l^2 [q_{1}^{\perp}\times q_{2}^{\perp}]
-2 l^2 q_{1}^{\perp}\cdot q_{2}^{\perp}\right]}
{\left((q_{1}^{\parallel})^2-m_{dyn}^2\right)
\left((q_{2}^{\parallel})^2-m_{dyn}^2\right)
\left[(q_{1}^{\parallel}+q_{2}^{\parallel})^2-m_{dyn}^2\right]^2}
\nonumber\\
&&\times  \gamma^{\sigma} (\hat{q}_2^{\parallel}+m_{dyn}) O^{(-)}
\gamma^{\nu} (\hat{q}_{1}^{\parallel}
+\hat{q}_2^{\parallel}+m_{dyn})
O^{(-)} \gamma^{\mu} (\hat{q}_{1}^{\parallel}
+\hat{q}_2^{\parallel}+m_{dyn}) O^{(-)}\nonumber\\
&&\times
\gamma^{\rho} (\hat{q}_1^{\parallel}+m_{dyn})
O^{(-)}\gamma^{\lambda}
{\cal D}_{\nu\lambda}(q_{1}^{\parallel},q_{1}^{\perp})
{\cal D}_{\sigma\rho}(q_{2}^{\parallel},q_{2}^{\perp}).
\end{eqnarray}
Switching to the Euclidean space and substituting the photon
propagator in the Feynman like gauge (\ref{non-l}), we find that
the only nonzero contribution is proportional to $O^{(-)}$,
\begin{eqnarray}
&&\tilde{\Gamma}^{(2)\mu}(0,0) =\frac{4\alpha^2 m_{dyn}^4}{\pi^6}
\gamma^{\mu}_{\parallel} O^{(-)} \nonumber\\
&&\times \int \frac{d^2 q_{1} d^2 q_{2}
d^2 q_{1}^{\perp} d^2 q_{2}^{\perp}
\exp\left[-\frac{3}{2}(q_{1}^{\perp}l)^2
-\frac{3}{2}(q_{2}^{\perp}l)^2
+2i l^2 [q_{1}^{\perp}\times q_{2}^{\perp}]
-2 l^2 q_{1}^{\perp}\cdot q_{2}^{\perp}\right]}
{\left(q_{1}^2+m_{dyn}^2\right)
\left(q_{2}^2+m_{dyn}^2\right)
\left[(q_{1}+q_{2})^2+m_{dyn}^2\right]^2
\left(q_{1}^2+(q_{1}^{\perp})^2+M_{\gamma}^2\right)
\left(q_{2}^2+(q_{2}^{\perp})^2+M_{\gamma}^2\right)}
\nonumber\\
&&\leq \mbox{Const}\gamma^{\mu}_{\parallel} O^{(-)} \alpha^{2}
\ln^2\left(\frac{1+(M_{\gamma}l)^2 }{(M_{\gamma}l)^2}\right).
\label{A8}
\end{eqnarray}
If we assume that $N$ is not too large, so that
$(M_{\gamma}l)^2\simeq 2\bar{\alpha}/\pi \ll 1$, then
the above estimate for the two-loop correction becomes
\begin{equation}
\tilde{\Gamma}^{(2)\mu}(0,0)\leq 
\mbox{Const}~\gamma^{\mu}_{\parallel}
O^{(-)} \alpha^{2} \ln^2\left(\frac{\pi}{2\bar{\alpha}}\right).
\label{A10}
\end{equation}
Thus, we conclude that, as expected, the Feynman like non-local
gauge leads to suppressed higher order corrections. The latter,
in its turn, means that the solution to the SD equation with the
bare vertex and the photon propagator in Eq.~(\ref{non-l}) is
reliable, and that the result presumably approaches the exact one
when the fine structure constant is very small.

At this point it is instructive to explicitly demonstrate that
the observed suppression in higher orders is the exclusive
property of the special gauge in Eq.~(\ref{non-l}).

To prove this, let us consider the one-loop vertex correction in
the case of the covariant gauge (\ref{cov-gauge}). After
substituting the photon propagator into Eq.~(\ref{A5}), we arrive
at the following estimate
\begin{eqnarray}
\tilde{\Gamma}^{(1)\mu}(0,0)&\simeq&\frac{\alpha}{4\pi}
\gamma_{\parallel}^{\mu}O^{(-)}
\ln^2\left(\frac{1}{(m_{dyn}l)^2}\right)
-\frac{\alpha}{4\pi} \gamma_{\parallel}^{\mu}O^{(-)}
\ln\left(\frac{M_{\gamma}^2}{m_{dyn}^2(m_{dyn}l)^2}\right)
\ln\left(\frac{1}{(M_{\gamma}l)^2}\right) \nonumber\\
&&-\lambda\frac{\alpha}{2\pi}
\gamma_{\parallel}^{\mu}O^{(-)}
\ln\left(\frac{1}{(m_{dyn}l)^2}\right),
\label{A17}
\end{eqnarray}
where, as is easy to check, the leading term with double
logarithm comes from the $q_{\mu}^{\parallel}
q_{\nu}^{\parallel}/q^2 q_{\parallel}^{2}$ component in the
photon propagator in Eq.~(\ref{cov-gauge}). Since in the rainbow
approximation the solution to the SD equation in the covariant
gauge yields $\ln^2(m_{dyn}l)^2\sim 1/\alpha$ [see
Eq.~(\ref{m_dyn})], we conclude that the one-loop correction in
Eq.~(\ref{A17}) is of the same order as the bare vertex, {\em
i.e.}, $\tilde{\Gamma}^{(1)\mu}(0,0)=O(1)$. As a result, such an
approximation is not self-consistent.

Notice that if $M_\gamma^2$ were equal to zero ({\em i.e.}, $
\Pi(q_\perp^2,q_\parallel^2)=0$ in Eq.~(\ref{cov-gauge}) and the
$q_\mu^\parallel q_\nu^\parallel/{q^2q_\parallel^2}$ term would
be absent there), the first two terms in Eq.~(\ref{A17}) would
cancel, and the one-loop correction
$\tilde{\Gamma}^{(1)\mu}(0,0)$ would be small
($\tilde{\Gamma}^{(1)\mu}(0,0)\sim O(\sqrt\alpha)$). As a result,
expression (\ref{m_dyn}) for $m_{dyn}$ would be correct.
Therefore the origin of the deviation of expressions
(\ref{m_dyn}) and (\ref{m}) for $m_{dyn}$ is the generation of a
non-zero $M_\gamma$.

We emphasize that the $q_{\mu}^{\parallel}
q_{\nu}^{\parallel}/q^2q_{\parallel}^{2}$ component in the
photon propagator, responsible for breaking the consistency of
the rainbow approximation to the SD equation, is absent in the
gauge Eq.~(\ref{non-l}). As a result, in this gauge, there exists
a consistent truncation of the SD equation for small values of
the coupling constant. Since the dynamical mass is a gauge
invariant quantity, in other gauges one needs to sum up an
infinite set of diagrams to recover the same result. Obviously
it is rather difficult to classify all the relevant diagrams in
those gauges, and so the existence of the special gauge is the
key point that allows to solve the problem of the magnetic
catalysis in QED reliably.

\section{Analytical analysis of the SD equation}
\label{secB}

In this appendix we will describe the analytical solution of the
SD equation (\ref{SD-B}).

First of all, let us show that the leading singularity,
$1/{\alpha\log\alpha}$, in $\log(m_{dyn}^2)$ in
Eq.~(\ref{dynmass}) is induced in the kinematic region with
$m_{\rm dyn}^2\ll|q_{\parallel}^2|\ll|eB|$ and $m_{dyn}^2\ll
M_\gamma^2\alt q_\perp^2\ll|eB|$ (in that region, fermions can
be treated as massless).

As was shown in Sec.~\ref{secV}, the approximation with
$B(p_\parallel^2)=m_{\rm dyn}$ for $p_\parallel^2<2|eB|$ and
$B(p_\parallel^2)$ rapidly decreasing for $p_\parallel^2>2|eB|$
is reliable in this problem (see Fig.~\ref{fig-b}). Then, taking
$p_\parallel^2=0$ in Eq.~(\ref{SD-B}), we arrive at the equation
\begin{eqnarray}
1&=&\frac{\alpha}{2\pi^2}\int^{2|eB|}\frac{d^2q_\parallel}
{q_\parallel^2+m_{dyn}^2} \int\limits_0^\infty
\frac{dx\exp\left(-xl^2/2\right)}{x+q_\parallel^2+q_\parallel^2
\Pi_E(x,q_\parallel^2)}\nonumber\\
&\simeq&\frac{\alpha}{2\pi^2}\int^{2|eB|}\frac{d^2q_\parallel}
{q_\parallel^2+m_{dyn}^2} \int\limits_0^{2|eB|}
\frac{dx}{x+q_\parallel^2+q_\parallel^2\Pi_E(x, q_\parallel^2)}.
\label{C1}
\end{eqnarray}
Matching now the asymptotes (\ref{Pi-IR}),(\ref{Pi-UV}) at
$q_\parallel^2=6m_{dyn}$ in Euclidean space, we get
\begin{eqnarray}
1&\simeq&\frac{\alpha}{2\pi}\int\limits_0^{2|eB|}dx\left[\int
\limits_0^{6m_{dyn}^2}\frac{dy}{(y+
m_{dyn}^2)\left(x+y(1+\frac{M_\gamma^2}{6 m_{\rm
dyn}^2})\right)}\right.\nonumber\\ &+&\left.\int\limits_{6m_{\rm
dyn}^2}^{2|eB|}\frac{dy}{(y+ m_{dyn}^2)\left(x+y+M_\gamma^2
e^{-xl^2/2}\right)}\right].
\label{C2}
\end{eqnarray}
It is clear that, because of $m_{dyn}^2$ in $(y+m_{dyn}^2)$, the
first term in the square bracket on the right hand side of this
equation is of order $O(1)$ and can be neglected: it cannot give
a contribution of order $1/{\alpha\log\alpha}$ to $m_{dyn}^2$.
Then we arrive at the estimate,
\begin{equation}
1\simeq\frac{\alpha}{2\pi}\int\limits_{6m_{\rm
dyn}^2}^{2|eB|}\frac{dy}{y+ m_{dyn}^2}
\int\limits_0^{2|eB|}\frac{dx}{x+y+M_\gamma^2e^{-xl^2/2}}.
\label{C3}
\end{equation}
The double logarithmic contribution comes from the region 
$2|eB|\gg y=q_{\parallel}^2\gg m_{dyn}^2$, $2|eB|\gg
x=q_{\perp}^2\agt y+ M_\gamma^2\geq M_\gamma^2,M_\gamma^2
=2\bar\alpha/{\pi l^2}$. Therefore one can write 
\begin{equation}
1\simeq\frac{\alpha}{2\pi}\int\limits_{6m_{dyn}^2}^{2|eB|}
\frac{dy}{y}
\int\limits_{y+M_\gamma^2}^{2|eB|}\frac{dx}{x}=
\frac{\alpha}{2\pi}\int\limits_{6m_{dyn}^2}^{2|eB|}
\frac{dy}{y}\log\frac{2|eB|}{y+M_\gamma^2}.
\label{C4}
\end{equation}
To calculate the last integral with  double logarithmic accuracy,
we write
\begin{eqnarray}
1&\simeq&\frac{\alpha}{2\pi}\left[\log\frac{2eB}{M_\gamma^2}\int
\limits_{6m_{dyn}^2}^{M_\gamma^2}\frac{dy}{y}
+\int\limits_{M_\gamma^2}
^{2|eB|}\frac{dy}{y}\log\frac{2|eB|}{y}\right]\nonumber\\
&\simeq&\frac{\alpha}{2\pi}
\left[\log\frac{2eB}{M_\gamma^2}\log\frac
{M_\gamma^2}{ m_{dyn}^2}+{1\over2}\log^2\frac{2eB}
{M_\gamma^2}\right]=
\frac{\alpha}{4\pi}\log\frac{2eB}{M_\gamma^2}
\log\left[\frac{2|eB| M_\gamma^2}{ m_{dyn}^4}\right].
\label{C5}
\end{eqnarray}
This equation implies that
\begin{equation}
m_{dyn}\sim\sqrt{|eB|}\left(\frac{N\alpha}{\pi}\right)^{1/4}
\exp\left(-\frac{\pi}{\alpha\log{\pi\over N\alpha}}\right).
\label{C6}
\end{equation}
Comparing this expression with Eq.~(\ref{dynmass}), one can see
that this estimate is quite reasonable. The origin of that is a
rather simple form of the fermion mass function
$B(p_\parallel^2)$: $B(p_\parallel^2)\simeq m_{dyn}$ for
$p_\parallel^2\alt 2|eB|$ and $B(p_\parallel^2)$ rapidly
decreases for $p_\parallel^2\agt 2|eB|$.

Therefore the dominant contribution to the SD equation
(\ref{SD-B}) comes from the region with $m_{\rm dyn}^2
\ll|q_{\parallel}^2| \ll|eB|$ and $m_{dyn}^2\ll M_\gamma^2\alt
q_\perp^2\ll|eB|$, where fermions can be treated as massless.
This in turn justifies the approximation with the polarization
function $\Pi_E=2\bar\alpha|eB|/{\pi q_\parallel^2}$.

Now we proceed at solving analytically SD equation (\ref{SD-B})
for a mass function in the improved ladder approximation. It is
\begin{equation}
B(p)=\frac{\alpha}{2\pi^2}\int\frac{d^2kB(k)}{k^2+B^2(k)}\int
\limits_0^\infty\frac{dze^{-zl^2/2}}{z+(k-p)^2+\Pi(z)},\quad
\Pi(z)=M_\gamma^2e^{-zl^2/2},
\label{masseq}
\end{equation}
where $M_\gamma^2=2\bar{\alpha} |eB|/\pi$ and we shifted the
momentum of integration (also we omit the symbol $\parallel$ in
the rest of this appendix). After the integration over the
angular coordinate, Eq.~(\ref{masseq}) becomes
\begin{equation}
B(p^2)=\frac{\alpha}{2\pi}\int\frac{dk^2B(k^2)}
{k^2+B^2(k^2)}K(p^2,k^2)
\label{scalareq}
\end{equation}
with the kernel
\begin{equation}
K(p^2,k^2)=\int\limits_0^\infty\frac{dz\exp(-zl^2/2)}
{\sqrt{(p^2+k^2+M_\gamma^2e^{-zl^2/2}+z)-4p^2k^2}}.
\end{equation}
To study Eq.~(\ref{scalareq}) analytically, we break up the
momentum integration into two regions and expand the kernel
appropriately for each region (compare with Refs.
\cite{Appelquist,3})
\begin{eqnarray}
B(p^2)&=&\frac{\alpha}{2\pi}
\left[\int\limits_0^{p^2}\frac{dk^2B(k^2)}{k^2+B^2(k^2)}
\int\limits_0^\infty\frac{dz\exp(-zl^2/2)}{p^2+M_\gamma^2
e^{-z l^2/2}+z}\right.\nonumber\\
&+&\left.\int\limits_{p^2}^\infty\frac{dk^2B(k^2)}{k^2+B^2(k^2)}
\int\limits_0^\infty\frac{dz\exp(-zl^2/2)}
{k^2+M_\gamma^2e^{-zl^2/2}+z} \right].
\label{approxeq}
\end{eqnarray}
Introducing dimensionless variables $x=p^2l^2/2,\,y=k^2l^2/2$ and
also the dimensionless mass function $B(p^2)/\sqrt{2|eB|} \to
B(x)$, we rewrite the last equation in the form
\begin{equation}
B(x)=\frac{\alpha}{2\pi}\left[g(x)\int\limits_0^x
\frac{dyB(y)}{y+B^2(y)}+\int\limits_x^\infty\frac{dyB(y)g(y)}
{y+B^2(y)}\right], \label{inteq} 
\end{equation}
where
\begin{equation}
g(x)=\int\limits_0^\infty\frac{dze^{-z}}
{z+x+\frac{\bar\alpha}{\pi}e^{-z}}.
\end{equation}
 The solutions of
the integral equation (\ref{inteq}) satisfy the second-order
differential equation
\begin{equation}
B^{\prime\prime}-\frac{g^{\prime\prime}}{g^\prime}B^\prime
-\frac{\alpha}{2\pi}g^\prime\frac{B}{x+B^2(x)}=0,
\label{diffeq}
\end{equation}
where the prime denotes derivative with respect to $x$. The
boundary conditions are
\begin{equation}
\frac{B^\prime}{g^\prime}\Bigg|_{x=0}=0,
\label{IRBC}
\end{equation}
\begin{equation}
\left(B-\frac{gB^\prime}{g^\prime}\right)\Bigg|_{x=\infty}=0.
\label{UVBC}
\end{equation}
The function $g(x)$ has asymptotic behavior
\begin{eqnarray}
g(x)&\sim&\log\frac{1+{\bar\alpha\over
\pi}}{x+{\bar\alpha\over\pi}},\qquad x\ll 1,\nonumber\\
g(x)&\sim&\frac{1}{x},\qquad x\gg 1.
\end{eqnarray}
We consider now the linearized version of Eq.~(\ref{diffeq}) when
the term $B^2(x)$ in denominator is replaced by a constant
$B^2(0) \equiv a^2$ ($B(p=0)\equiv m_{dyn}$): the numerical
analysis shows that it is an excellent approximation. The two
independent solutions of that equation near the point $x=\infty$
behave as $B(x)\sim {\rm const}$ and $B(x)\sim 1/x$, and the UVBC
selects the last one.

In the region $x\ll 1$, the equation takes the form
\begin{equation}
B^{\prime\prime}+\frac{1}{x+{\bar\alpha\over\pi}}
B^\prime+\frac{\alpha}
{2\pi}\frac{B}{(x+{\bar\alpha\over\pi})(x+a^2)}=0.
\label{lineareq}
\end{equation}
Introducing the variable $x+a^2=-z(\bar\alpha/\pi -a^2)$,
Eq.~(\ref{lineareq}) can be rewritten in the form of an equation
for the hypergeometric function,
\begin{equation}
z(1-z)\frac{d^2B}{dz^2}-z\frac{dB}{dz}-\frac{\alpha}{2\pi}B=0.
\label{hyper}
\end{equation}
The general solution to Eq.~(\ref{hyper}) has the form
\begin{equation}
B(z)=C_1u_1+C_2u_2,
\label{genersol}
\end{equation}
where
\begin{equation}
u_1=zF(1+i\nu,1-i\nu;2;z),
\end{equation}
\begin{equation}
u_2=(-z)^{-i\nu}F(i\nu,1+i\nu;1+2i\nu;{1\over z})+(-z)^{i\nu}
F(-i\nu,1-i\nu;1-2i\nu;{1\over z}),
\end{equation}
$\nu=\sqrt{\alpha/2\pi}$. From the infrared boundary condition
(\ref{IRBC}), one gets
\begin{equation}
\frac{C_2}{C_1}=-\frac{u_1^\prime}{u_2^\prime}\Bigg|_{x=0}.
\end{equation}
Equating now the logarithmic derivatives of solution
(\ref{genersol}) (at $x\ll 1$) and $1/x$ (at $x\gg 1$) at the
point $x=1$, we arrive at the equation determining the quantity
$a(\alpha)$ ({\em i.e.}, the dynamical mass $m_{dyn}$):
\begin{equation}
\varphi\equiv A_1B_2-A_2B_1=0,
\label{phieq}
\end{equation}
where
\begin{equation}
A_i=\left(u_i^\prime+u_i\right)\Bigg|_{x=1},\qquad
B_i=u_i^\prime\Bigg|_{x=0}.
\end{equation}
Since the variable $z$ is
\begin{equation}
z=-\frac{x+a^2}{{\bar\alpha\over\pi}-a^2}\Bigg|_{x=0}\simeq
-\frac{\pi}{\bar\alpha}a^2,\quad z
=-\frac{x+a^2}{{\bar\alpha\over\pi}-a^2}
\Bigg|_{x=1}\simeq -\frac{\pi}{\bar\alpha}
\end{equation}
(we suppose $a^2\ll\bar\alpha/\pi$), in what follows we need
asymptotic behavior of $u_i(z),u_i^\prime (z)$ at small  and
large negative values of its argument $z$. Using corresponding
formulas from \cite{Bateman}, we find for small values of $z$
($|z|\ll 1$) 
\begin{eqnarray}
u_1&\simeq& z(1+\frac{1+\nu^2}{2}z)+ O(z^2)\\
 u_2&\simeq&2Re\left\{\frac{\Gamma(1+2i\nu)}{\Gamma^2(1+i\nu)}
 \left[ \nu^2z(\ln(-z) -h_0)+1\right]+O(z^2\ln z)\right\},\\
  u_1^\prime &\simeq& 1+(1+\nu^2) z +O(z^2), \\
u_2^\prime&\simeq&2Re\left\{\frac{\Gamma(1+2i\nu)}
{\Gamma^2(1+i\nu)}
 \nu^2(\ln(-z)+1 -h_0)+O(z\ln z)\right\},
\end{eqnarray}
where
\begin{equation}
h_0=1-2\gamma-\psi(i\nu) -\psi(1+i\nu).
\end{equation}
 At large $|z|\gg 1$ we have
\begin{eqnarray}
u_1 &\simeq& -\frac{1}{\nu}\sqrt{\frac{\tanh(\pi\nu)}{\pi\nu}}
\sin\left[\nu\ln(-z)+\Phi(\nu)\right]+O(z^{-1}),\\
u_2 &\simeq& 2 \cos\left(\nu\ln (-z)\right) +O(z^{-1}),\\
u_1^\prime &\simeq& 
-\frac{1}{z}\sqrt{\frac{\tanh(\pi\nu)}{\pi\nu}}
\cos\left(\nu\ln(-z)+\Phi(\nu)\right)+O(z^{-2}), \\
u_2^\prime &\simeq& -\frac{2\nu}{z} \sin\left(\nu\ln(-z)\right)
+O(z^{-2}),
\end{eqnarray}
where
\begin{eqnarray}
\Phi(\nu)&=&\mbox{arg}
\left(\frac{\Gamma(1+2i\nu)}{\Gamma^2(1+i\nu)}\right)
=\sum_{n=1}^{\infty}(-1)^{n+1}\frac{2(2^{2n}-1)\zeta(2n+1)}
{2n+1}\nu^{2n+1} \\ &\simeq& 2
\zeta(3)\nu^{3}-6\zeta(5)\nu^{5}+\dots.
\end{eqnarray}
By making use of these asymptotes, we obtain the following
expressions for $A_i,B_i$,
\begin{eqnarray}
A_1 &=& (\frac{du_1}{dx}+u_1)|_{x=1}\simeq
-\sqrt{\frac{\tanh(\pi\nu)}{\pi\nu}}
\left[\cos(\nu\ln\frac{\pi}{\bar\alpha})
+\frac{1}{\nu}\sin\left(\nu\ln\frac{\pi}
{\bar\alpha}+\Phi(\nu)\right)\right],
\\ A_2 &=&(\frac{du_2}{dx}+ u_2)|_{x=1}
\simeq2 \cos\left(\nu\ln\frac{\pi}
{\bar\alpha}\right)+2\sin\left(\nu\ln
\frac{\pi}{\bar\alpha}\right),\\
 B_1 &=& \frac{du_1}{dx} |_{x=0} 
\simeq -\frac{\pi}{\bar\alpha}, \\
B_2&=& \frac{du_2}{dx} |_{x=0} \simeq -\ln\frac{\pi
a^2}{\bar\alpha}.
\end{eqnarray}

And, finally, the solution to Eq.~(\ref{phieq}) reads
\begin{equation}
a^2=\frac{m_{dyn}^2}{2|eB|}\simeq
\frac{N\alpha}{\pi}\exp\left[-\frac{1}{\nu}
\cot\left(\nu\ln\frac{\pi}{N\alpha}\right)\right]
\simeq\left(N\alpha \over\pi\right)^{2/3}
\exp\left[-\frac{2\pi}{\alpha\ln (\pi/N\alpha)}\right],
\quad \mbox{as} \quad \alpha\to 0.
\label{analdynmass}
\end{equation}
The obtained analytical expression (\ref{analdynmass}) for the
dynamical mass is close both to estimate (\ref{C6}) and to the
numerical solution (\ref{dynmass}). The ratio of the values of
$C_1$ in the analytical solution and in the numerical one is
$C_1^{(\rm anal)}/C_1^{(\rm numer)}\simeq 1.7$. This, rather
mild, discrepancy reflects the approximations made in the kernel
of the integral equation (\ref{scalareq}) in reducing it to the
differential equation (\ref{diffeq}).

\section{Ward-Takahashi identity}
\label{secC}

In this appendix, we discuss the simplest Ward-Takahashi (WT)
identity relating the fermion propagator and the vertex in the
problem of magnetic catalysis (for another approach see 
Ref.~\cite{20}).

Let us start from the WT identity in coordinate space:
\begin{equation}
\partial_{\mu}^{z}\Gamma^{\mu}(x',y';z)
=\delta(x'-z)G^{-1}(z,y')-\delta(z-y')G^{-1}(x',z),
\end{equation}
After multiplying this expression with $G(x,x')$ on the left and
$G(y',y)$ on the right and integrating over $x'$ and $y'$, we
arrive at another representation,
\begin{equation}
\int d^4 x' d^4 y' G(x,x') \partial_{\mu}^{z}
\Gamma^{\mu}(x',y';z)
G(y',y)=\delta(z-y)G(x,z)-\delta(x-z)G(z,y).
\label{wt1}
\end{equation}
By taking into account the universal Schwinger phase factors of
the propagators and the vertex [see Eq.~(\ref{14}) in
Sec.~\ref{secII}], this identity (in the momentum representation)
reads 
\begin{eqnarray}
i \tilde{G}(p_2) -i \tilde{G}(p_1)&=&
\int d^4 r \frac{d^4 q}{(2\pi)^4} e^{-irq}
\tilde{G}\left(p_1+A^{ext}(r)\right) (p_1-p_2)_{\mu} 
\tilde{\Gamma}^{\mu}(p_1+q,p_2+q)
\tilde{G}\left(p_2+q-A^{ext}(r)\right). 
\end{eqnarray}
Now, by substituting the bare vertex along with the solution
for the fermion propagator as in Eq.~(\ref{gen-sol}) with
$A(p_{\parallel}^2)=1$ and $B(p_{\parallel}^2)=m_{dyn}=Const$
(recall that the latter is a very good approximation in the most
important region of momenta, $p_{\perp,\parallel}^2\ll |eB|$), we
obtain the following relation
\begin{eqnarray}
&&\left[ (\hat{p}_1^{\parallel}-m_{dyn}) e^{-(p_2^{\perp} l)^2} 
-(\hat{p}_2^{\parallel}-m_{dyn}) e^{-(p_1^{\perp} l)^2} \right] 
O^{(-)} = e^{-(p_1^{\perp} l)^2/2-(p_2^{\perp} l)^2/2
+i l^2 [p_1^{\perp}\times p_2^{\perp}]}
(\hat{p}_1^{\parallel}-\hat{p}_2^{\parallel}) O^{(-)}.
\label{wt-id}
\end{eqnarray}
Once again, the approximation for the fermion propagator with
$A(p_{\parallel}^2)=1$ and $B(p_{\parallel}^2)=m_{dyn}=Const$ is
reliable only in the dynamical region $p_{\perp,\parallel}^2\ll
|eB|$. Therefore, at best, one could expect that the WT identity
is satisfied in this same region of momenta. After expanding 
both sides of Eq.~(\ref{wt-id}) in powers of $p_{\perp}^2 l^2
=p_{\perp}^2/|eB|$ and keeping only the leading order
contribution, we see that Eq.~(\ref{wt-id}) indeed turns into an
identity. In the end, we remind that the same conclusion about
the WT identity was  reached in Ref.~\cite{20} by making use of a
different approach.

\newpage
%%%%%%%%%%%%%%%%%%%%% Fig.1 %%%%%%%%%%%%%%%%%%%%%
%%%%%%%%%%%%%%%%%%%%% Fig.2 %%%%%%%%%%%%%%%%%%%%%
%%%%%%%%%%%%%%%%%%%%% Fig.3 %%%%%%%%%%%%%%%%%%%%%
%%%%%%%%%%%%%%%%%%%%% Fig.4 %%%%%%%%%%%%%%%%%%%%%
%%%%%%%%%%%%%%%%%%%%% Fig.5 %%%%%%%%%%%%%%%%%%%%%
%%%%%%%%%%%%%%%%%%%%% End %%%%%%%%%%%%%%%%%%%%%

\end{document}